\documentclass[pdftex,twocolumn,epjc3]{svjour3}          

\RequirePackage{amsmath}
\RequirePackage{amssymb}
\RequirePackage{multirow}
\RequirePackage{subfigure}
\RequirePackage{hyperref}
\RequirePackage{widetext}

\allowdisplaybreaks[3]

\RequirePackage[T1]{fontenc}

\smartqed  

\RequirePackage{graphicx}

\journalname{Eur. Phys. J. C}

\begin{document}

\title{Decay properties of $P_c$ states through the Fierz rearrangement}

\author{Hua-Xing Chen}

\institute{School of Physics, Southeast University, Nanjing 210094, China}

\date{Received: date / Accepted: date}

\maketitle

\begin{abstract}
We systematically study hidden-charm pentaquark currents with the quark configurations $[\bar c u][u d c]$, $[\bar c d][u u c]$, and $[\bar c c][u u d]$. Some of their relations are derived using the Fierz rearrangement of the Dirac and color indices, and the obtained results are used to study strong decay properties of $P_c$ states as $\bar D^{(*)} \Sigma_c$ hadronic molecules. We calculate their relative branching ratios for the $J/\psi p$, $\eta_c p$, $\chi_{c0} p$, $\chi_{c1} p$, $\bar D^{(*)0} \Lambda_c^+$, $\bar D^{0} \Sigma_c^{+}$, and $\bar D^{-} \Sigma_c^{++}$ decay channels. We propose to search for the $P_c(4312)$ in the $\eta_c p$ channel and the $P_c(4440)/P_c(4457)$ in the $\bar D^{0} \Lambda_c^+$ channel.
\end{abstract}

\section{Introduction}
\label{sec:intro}

Since the discovery of the $X(3872)$ in 2003 by Belle~\cite{Choi:2003ue}, many charmonium-like $XYZ$ states were discovered in the past twenty years~\cite{pdg}. Besides, the LHCb Collaboration observed three enhancements in the $J/\psi p$ invariant mass spectrum of the $\Lambda_b\to J/\psi p K$ decays~\cite{Aaij:2015tga,Aaij:2019vzc}:
\begin{eqnarray}
\nonumber     P_c(4312)^+:    &M=& 4311.9 \pm  0.7 ^{+6.8}_{-0.6}  \mbox{ MeV} \, ,
\\ \nonumber                  &\Gamma=&    9.8 \pm  2.7 ^{+3.7}_{-4.5}  \mbox{ MeV} \, ,
\\[1mm]            P_c(4440)^+:    &M=& 4440.3 \pm  1.3 ^{+4.1}_{-4.7}  \mbox{ MeV} \, ,
\label{experiment}
\\ \nonumber                  &\Gamma=&   20.6 \pm  4.9 ^{+8.7}_{-10.1} \mbox{ MeV} \, ,
\\[1mm] \nonumber  P_c(4457)^+:    &M=& 4457.3 \pm  0.6 ^{+4.1}_{-1.7}  \mbox{ MeV} \, ,
\\ \nonumber                  &\Gamma=&    6.4 \pm  2.0 ^{+5.7}_{-1.9}  \mbox{ MeV} \, .
\end{eqnarray}
These structures contain at least five quarks, $\bar c c u u d$, so they are perfect candidates of hidden-charm pentaquark states. Together with the charmonium-like $XYZ$ states, their studies are significantly improving our understanding of the non-perturbative behaviors of the strong interaction at the low energy region~\cite{Chen:2016qju,Liu:2019zoy,Lebed:2016hpi,Esposito:2016noz,Guo:2017jvc,Ali:2017jda,Olsen:2017bmm,Karliner:2017qhf,Brambilla:2019esw,Guo:2019twa}.

The $P_c(4312)$, $P_c(4440)$, and $P_c(4457)$ are just below the $\bar D \Sigma_c$ and $\bar D^{*} \Sigma_c$ thresholds, so it is quite natural to interpret them as $\bar D^{(*)} \Sigma_c$ hadronic molecular states, whose existence had been predicted in Refs.~\cite{Wu:2010jy,Wang:2011rga,Yang:2011wz,Karliner:2015ina,Wu:2012md} before the LHCb experiment performed in 2015~\cite{Aaij:2015tga}. This experiment observed two structures $P_c(4380)$ and $P_c(4450)$. Later in 2019 another LHCb experiment~\cite{Aaij:2019vzc} observed a new structure $P_c(4312)$ and further separated the $P_c(4450)$ into two substructures $P_c(4440)$ and $P_c(4457)$.

To explain these $P_c$ states, various theoretical interpretations were proposed, such as loosely-bound meson-baryon molecular states~\cite{Chen:2019asm,Liu:2019tjn,He:2019ify,Huang:2019jlf,Guo:2019kdc,Fernandez-Ramirez:2019koa,Xiao:2019aya,Meng:2019ilv,Wu:2019adv,Wang:2019hyc,Yamaguchi:2019seo,Valderrama:2019chc,Liu:2019zvb,Burns:2019iih,Wang:2019ato,Gutsche:2019mkg,Du:2019pij,Azizi:2016dhy,Chen:2019bip} and tightly-bound pentaquark states~\cite{Maiani:2015vwa,Lebed:2015tna,Stancu:2019qga,Giron:2019bcs,Ali:2019npk,Weng:2019ynv,Eides:2019tgv,Wang:2019got,Cheng:2019obk,Ali:2019clg}, etc. Since they have only been observed in the $J/\psi p$ invariant mass spectrum by LHCb~\cite{Aaij:2015tga,Aaij:2019vzc}, it is crucial to search for some other decay channels in order to better understand their nature. There have been some theoretical studies on this subject, using the heavy quark symmetry~\cite{Voloshin:2019aut,Sakai:2019qph}, effective approaches~\cite{Guo:2019fdo,Xiao:2019mst,Cao:2019kst,Lin:2019qiv}, QCD sum rules~\cite{Xu:2019zme}, and the quark interchange model~\cite{Wang:2019spc}, etc. We refer to reviews~\cite{Chen:2016qju,Liu:2019zoy,Lebed:2016hpi,Esposito:2016noz,Guo:2017jvc,Ali:2017jda,Olsen:2017bmm,Karliner:2017qhf,Brambilla:2019esw,Guo:2019twa} and references therein for detailed discussions.

In this paper we shall apply the Fierz rearrangement of the Dirac and color indices to study strong decay properties of $P_c$ states as $\bar D^{(*)} \Sigma_c$ hadronic molecules, which method has been used in Ref.~\cite{Chen:2019wjd} to study strong decay properties of the $Z_c(3900)$ and $X(3872)$. A similar arrangement of the spin and color indices in the nonrelativistic case was used to study decay properties of $XYZ$ and $P_c$ states in Refs.~\cite{Voloshin:2013dpa,Maiani:2017kyi,Voloshin:2018pqn,Voloshin:2019aut,Wang:2019spc,Xiao:2019spy,Cheng:2020nho}.

In this paper we shall use the $\bar c$, $c$, $u$, $u$, and $d$ ($q=u/d$) quarks to construct hidden-charm pentaquark currents with the three configurations: $[\bar c u][u d c]$, $[\bar c d][u u c]$, and $[\bar c c][u u d]$. In Refs.~\cite{Chen:2015moa,Chen:2016otp,Xiang:2017byz} we have found that these three configurations can be related as a whole, while in the present study we shall further find that two of them are already enough to be related to each other, just with the color-octet-color-octet meson-baryon terms included. Using these relations, we shall study strong decay properties of $P_c$ states as $\bar D^{(*)} \Sigma_c$ molecular states.

Our strategy is quite straightforward. First we need a hidden-charm pentaquark current, such as
\begin{eqnarray}
\eta_1(x,y) &=& [\delta^{ab} \bar c_a(x) \gamma_5 u_b(x)]
\\ \nonumber && ~~~~~ \times [\epsilon^{cde} u_c^T(y) \mathbb{C} \gamma_\mu d_d(y) \gamma^\mu \gamma_5 c_e(y)] \, ,
\end{eqnarray}
where $a \cdots e$ are color indices. It is the current best coupling to the $\bar D^0 \Sigma_c^{+}$ molecular state of $J^P = 1/2^-$, through
\begin{equation}
\langle 0 | \eta_1(x,y) | \bar D^0 \Sigma_c^{+}; 1/2^-(q) \rangle = f_{P_c} u(q) \, ,
\end{equation}
where $u(q)$ is the Dirac spinor of the $P_c$ state.

After the Fierz rearrangement of the Dirac and color indices, we can transform it to be
\begin{eqnarray}
\eta_1(x,y) &\rightarrow& - {1\over12} ~ [\bar c_a(x^\prime) \gamma_5 c_a(x^\prime)]~N(y^\prime)
\\ \nonumber && ~~ + {1\over24} ~ [\bar c_a(x^\prime) \gamma_\mu c_a(x^\prime)]~\gamma^\mu \gamma_5 N(y^\prime) + \cdots \, ,
\end{eqnarray}
where
\begin{equation}
N = \epsilon^{abc} (u_a^T \mathbb{C} d_b) \gamma_5 u_c - \epsilon^{abc} (u_a^T \mathbb{C} \gamma_5 d_b) u_c \, ,
\end{equation}
is the Ioffe's light baryon field well coupling to the proton~\cite{Ioffe:1981kw,Ioffe:1982ce,Espriu:1983hu}. Hence, $\eta_1(x^\prime,y^\prime)$ couples to the $\eta_c p$ and $J/\psi p$ channels simultaneously:
\begin{eqnarray}
\langle 0 | \eta_1(x^\prime,y^\prime) | \eta_c p \rangle &\approx& - {1\over12} \langle 0 | \bar c_{a} \gamma_5 c_a | \eta_c \rangle~\langle 0 | N | p \rangle + \cdots \, ,
\\ \nonumber \langle 0 | \eta_1(x^\prime,y^\prime) | \psi p \rangle &\approx& {1\over24} \langle 0 | \bar c_{a} \gamma_\mu c_a | \psi \rangle~\gamma^\mu \gamma_5 \langle 0 | N | p \rangle + \cdots \, .
\end{eqnarray}
The above two equations can be easily used to calculate the relative branching ratio of the $P_c$ decay into $\eta_c p$ to its decay into $J/\psi p$~\cite{Yu:2017zst}. Detailed discussions on this will be given below.

This paper is organized as follows. In Sec.~\ref{sec:current} we systematically study hidden-charm pentaquark currents with the quark content $\bar c c u u d$. We consider three different configurations, $[\bar c u][u d c]$, $[\bar c d][u u c]$, and $[\bar c c][u u d]$, whose relations are derived in Sec.~\ref{sec:fierz} using the Fierz rearrangement of the Dirac and color indices. In Sec.~\ref{sec:decay} we extract some strong decay properties of $\bar D^{(*)0} \Sigma_c^+$ and $\bar D^{(*)-} \Sigma_c^{++}$ molecular states, which are combined in Sec.~\ref{sec:isospin} to further study strong decay properties of $\bar D^{(*)} \Sigma_c$ molecular states with $I = 1/2$. The results are summarized in Sec.~\ref{sec:summary}.

\section{Hidden-charm pentaquark currents}
\label{sec:current}

%
\begin{figure*}[hbt]
\begin{center}
\subfigure[~\mbox{$[\bar c u][u d c]$ current $\eta(x,y)$}]{\includegraphics[width=0.3\textwidth]{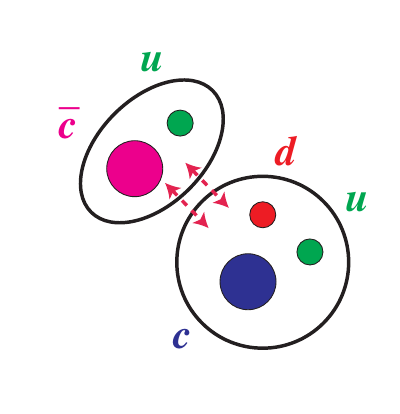}}
~~~~~~
\subfigure[~\mbox{$[\bar c d][u u c]$ current $\xi(x,y)$}]{\includegraphics[width=0.3\textwidth]{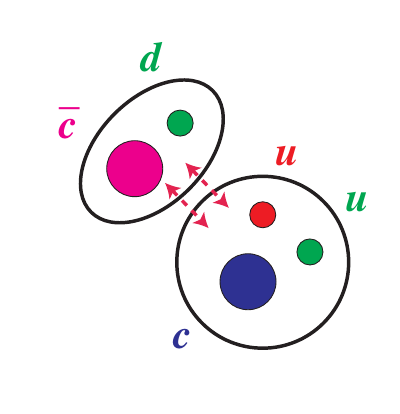}}
~~~~~~
\subfigure[~\mbox{$[\bar c c][u u d]$ current $\theta(x,y)$}]{\includegraphics[width=0.3\textwidth]{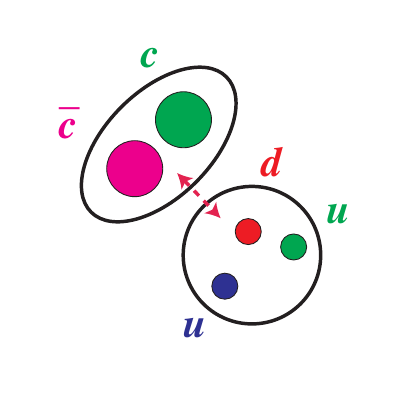}}
\caption{Three types of hidden-charm pentaquark currents. Quarks are shown in red/green/blue color, and antiquarks are shown in cyan/magenta/yellow color.}
\label{fig:current}
\end{center}
\end{figure*}
%

We can use $\bar c$, $c$, $u$, $u$, and $d$ ($q=u/d$) quarks to construct many types of hidden-charm pentaquark currents. In the present study we need the following three, as illustrated in Fig.~\ref{fig:current}:
\begin{eqnarray}
\nonumber    \eta(x,y)   &=& [\bar c_a(x) \Gamma^\eta_1 u_b(x)]    ~ \Big[[u^T_c(y) \mathbb{C} \Gamma^\eta_2 d_d(y)]    ~ \Gamma^\eta_3 c_e(y)   \Big]   \, ,
\\ \nonumber \xi(x,y)    &=& [\bar c_a(x) \Gamma^\xi_1 d_b(x)]   ~ \Big[[u^T_c(y) \mathbb{C} \Gamma^\xi_2 u_d(y)]   ~ \Gamma^\xi_3 c_e(y)  \Big]    \, ,
\\ \nonumber \theta(x,y) &=& [\bar c_a(x) \Gamma^\theta_1 c_b(x)] ~ \Big[[q^T_c(y) \mathbb{C} \Gamma^\theta_2 q_d(y)] ~ \Gamma^\theta_3 q_e(y)\Big] \, ,
\\
\end{eqnarray}
where $\Gamma_{1/2/3}^{\eta/\xi/\theta}$ are Dirac matrices, the subscripts $a \cdots e$ are color indices, and the sum over repeated indices (both superscripts and subscripts) is taken.

All the independent hidden-charm tetraquark currents of $J^{PC} = 1^{+\pm}$ have been constructed in Refs.~\cite{Chen:2008qw,Chen:2013jra,Chen:2010ze,Chen:2019wjd}. However, in this case there are hundreds of hidden-charm pentaquark currents, and it is difficult to find out all the independent ones (see Refs.~\cite{Chen:2015moa,Chen:2016otp} for relevant discussions). Hence, in this paper we shall not construct all the currents, but just investigate those that are needed to study decay properties of the $P_c(4312)$, $P_c(4440)$, and $P_c(4457)$. We shall separately investigate their color and Lorentz structures in the following subsections.

\subsection{Color structure}
\label{sec:color}

Taking $\eta(x,y)$ as an example, there are two possibilities to compose a color-singlet field: $[\bar c u]_{\mathbf{1}_c} [u  d  c]_{\mathbf{1}_c}$ and $[\bar c u]_{\mathbf{8}_c} [u  d  c]_{\mathbf{8}_c}$. We can use the color-singlet-color-singlet meson-baryon term
\begin{equation}
[\delta^{ab}\bar c_a u_b] [\epsilon^{cde} u_c d_d c_e] \, ,
\end{equation}
to describe the former, while there are three color-octet-color-octet meson-baryon terms for the latter:
\begin{eqnarray}
\nonumber    && [\lambda^{ab}_n\bar c_a u_b] [\epsilon^{cdf} \lambda^{fe}_n u_c d_d c_e] \, ,
\\           && [\lambda^{ab}_n\bar c_a u_b] [\epsilon^{def} \lambda^{fc}_n u_c d_d c_e] \, ,
\\ \nonumber && [\lambda^{ab}_n\bar c_a u_b] [\epsilon^{ecf} \lambda^{fd}_n u_c d_d c_e] \, .
\end{eqnarray}
Only two of them are independent due to
\begin{equation}
\epsilon^{cdf} \lambda^{fe}_n + \epsilon^{def} \lambda^{fc}_n + \epsilon^{ecf} \lambda^{fd}_n = 0 \, ,
\end{equation}
which is consistent with the group theory that there are two and only two octets in $\mathbf{3}_c \otimes \mathbf{3}_c \otimes \mathbf{3}_c = \mathbf{1}_c \oplus \mathbf{8}_c \oplus \mathbf{8}_c \oplus \mathbf{10}_c$. Similar argument applies to $\xi(x,y)$ and $\theta(x,y)$.

In Refs.~\cite{Chen:2015moa,Chen:2016otp} we use the color rearrangement
\begin{equation}
\delta^{ab} \epsilon^{cde} = \delta^{ac} \epsilon^{bde} + \delta^{ad} \epsilon^{cbe} + \delta^{ae} \epsilon^{cdb} \, ,
\end{equation}
together with the Fierz rearrangement to derive the relations among all the three types of currents, {\it e.g.}, we can transform an $\eta$ current into the combination of many $\xi$ and $\theta$ currents:
\begin{equation}
\eta \rightarrow \xi + \theta \, .
\end{equation}

In the present study we further derive another color rearrangement:
\begin{equation}
\delta^{ab} \epsilon^{cde} = {1\over3}~\delta^{ae} \epsilon^{bcd} - {1\over2}~\lambda^{ae}_n \epsilon^{bcf} \lambda^{fd}_n + {1\over2}~\lambda^{ae}_n \epsilon^{bdf} \lambda^{fc}_n \, .
\label{eq:color}
\end{equation}
{\it Note that the other color-octet-color-octet meson-baryon term $\lambda^{ae}_n \epsilon^{cdf} \lambda^{fb}_n$ can also be included, but the first coefficient $1/3$ always remains the same. This is reasonable because the probability of the relevant fall-apart decay is just 33\% if only considering the color degree of freedom, as shown in Figs.~\ref{fig:decayeta}(a) and \ref{fig:decayxi}(a).}

Using the above color rearrangement in the color space, together with the Fierz rearrangement in the Lorentz space to interchange the $u_b$ and $c_e$ quark fields, we can transform an $\eta$ current into the combination of many $\theta$ currents (both color-singlet-color-singlet and color-octet-color-octet ones). Similar arguments can be applied to relate
\begin{equation}
\eta \leftrightarrow \xi \, , ~~~ \xi \leftrightarrow \theta \, , ~~~ \theta \leftrightarrow \eta \, ,
\end{equation}
whose explicit formulae will be given in Sec.~\ref{sec:fierz}.

\subsection{$\eta/\xi(x,y)$ and heavy baryon fields}
\label{sec:current1}

\begin{table*}[hbt]
\begin{center}
\renewcommand{\arraystretch}{1.5}
\caption{Couplings of meson operators to meson states, where color indices are omitted for simplicity. Taken from Ref.~\cite{Chen:2019wjd}.}
\begin{tabular}{ c | c | c | c | c | c}
\hline\hline
~~Operators~~ & ~$I^GJ^{PC}$~ & ~~~~~~Mesons~~~~~~ & ~$I^GJ^{PC}$~ & ~~~Couplings~~~ & ~~~~~~Decay Constants~~~~~~
\\ \hline\hline
$I^{S} = \bar c c$                & $0^+0^{++}$                  & $\chi_{c0}(1P)$      & $0^+0^{++}$ & $\langle 0 | I^S | \chi_{c0} \rangle = m_{\chi_{c0}} f_{\chi_{c0}}$          & $f_{\chi_{c0}} = 343$~MeV~\mbox{\cite{Veliev:2010gb}}
\\ \hline
$I^{P} = \bar c i\gamma_5 c$      & $0^+0^{-+}$                  & $\eta_c$             & $0^+0^{-+}$ & $\langle 0 | I^{P} | \eta_c \rangle = \lambda_{\eta_c}$                      & $\lambda_{\eta_c} = {f_{\eta_c} m_{\eta_c}^2 \over 2 m_c}$
\\ \hline
$I^{V}_\mu = \bar c \gamma_\mu c$ & $0^-1^{--}$                  & $J/\psi$             & $0^-1^{--}$ & $\langle0| I^{V}_\mu | J/\psi \rangle = m_{J/\psi} f_{J/\psi} \epsilon_\mu$  & $f_{J/\psi} = 418$~MeV~\mbox{\cite{Becirevic:2013bsa}}
\\ \hline
\multirow{2}{*}{$I^{A}_\mu = \bar c \gamma_\mu \gamma_5 c$}
                               & \multirow{2}{*}{$0^+1^{++}$}    & $\eta_c$             & $0^+0^{-+}$ & $\langle 0 | I^{A}_\mu | \eta_c \rangle = i p_\mu f_{\eta_c}$                & $f_{\eta_c} = 387$~MeV~\mbox{\cite{Becirevic:2013bsa}}
\\ \cline{3-6}
                               &                              & $\chi_{c1}(1P)$      & $0^+1^{++}$ & $\langle 0 | I^{A}_\mu | \chi_{c1} \rangle = m_{\chi_{c1}} f_{\chi_{c1}} \epsilon_\mu $
                               &  $f_{\chi_{c1}} = 335$~MeV~\mbox{\cite{Novikov:1977dq}}
\\ \hline
\multirow{2}{*}{$I^{T}_{\mu\nu} = \bar c \sigma_{\mu\nu} c$}
                               & \multirow{2}{*}{$0^-1^{\pm-}$}  & $J/\psi$             & $0^-1^{--}$ & $\langle 0 | I^{T}_{\mu\nu} | J/\psi \rangle = i f^T_{J/\psi} (p_\mu\epsilon_\nu - p_\nu\epsilon_\mu) $
                               &  $f_{J/\psi}^T = 410$~MeV~\mbox{\cite{Becirevic:2013bsa}}
\\ \cline{3-6}
                               &                              & $h_c(1P)$            & $0^-1^{+-}$ & $\langle 0 | I^{T}_{\mu\nu} | h_c \rangle = i f^T_{h_c} \epsilon_{\mu\nu\alpha\beta} \epsilon^\alpha p^\beta $
                               &  $f_{h_c}^T = 235$~MeV~\mbox{\cite{Becirevic:2013bsa}}
\\ \hline\hline
$O^{S} = \bar c q$                & $0^{+}$                   & $\bar D_0^{*}$           & $0^{+}$  & $\langle 0 | O^{S} | \bar D_0^{*} \rangle = m_{D_0^{*}} f_{D_0^{*}}$             &  $f_{D_0^{*}} = 410$~MeV~\mbox{\cite{Narison:2015nxh}}
\\ \hline
$O^{P} = \bar c i\gamma_5 q$      & $0^{-}$                   & $\bar D$                & $0^{-}$  & $\langle 0 | O^{P} | \bar D \rangle = \lambda_D$                                &  $\lambda_D = {f_D m_D^2 \over {m_c + m_d}}$
\\ \hline
$O^{V}_\mu = \bar c \gamma_\mu q$ & $1^{-}$                   & $\bar D^{*}$        & $1^{-}$  & $\langle0| O^{V}_\mu | \bar D^{*} \rangle = m_{D^*} f_{D^*} \epsilon_\mu$   &  $f_{D^*} = 253$~MeV~\mbox{\cite{Chang:2018aut}}
\\ \hline
\multirow{2}{*}{$O^{A}_\mu = \bar c \gamma_\mu \gamma_5 q$}
                               & \multirow{2}{*}{$1^{+}$}      & $\bar D$         & $0^{-}$  & $\langle 0 | O^{A}_\mu | \bar D \rangle = i p_\mu f_{D}$                 &  $f_{D} = 211.9$~MeV~\mbox{\cite{pdg}}
\\ \cline{3-6}
                                  &                            & $\bar D_1$                & $1^{+}$  & $\langle 0 | O^{A}_\mu | \bar D_1 \rangle = m_{D_1} f_{D_1} \epsilon_\mu $        &  $f_{D_1} = 356$~MeV~\mbox{\cite{Narison:2015nxh}}
\\ \hline
\multirow{2}{*}{$O^{T}_{\mu\nu} = \bar c \sigma_{\mu\nu} q$}
                               & \multirow{2}{*}{$1^{\pm}$}    & $\bar D^{*}$        & $1^{-}$  &  $\langle 0 | O^{T}_{\mu\nu} | \bar D^{*} \rangle = i f_{D^*}^T (p_\mu\epsilon_\nu - p_\nu\epsilon_\mu) $
                               &  $f_{D^*}^T \approx 220$~MeV
\\ \cline{3-6}
                               &                               &  --                  & $1^{+}$  &  --  &  --
\\ \hline\hline
\end{tabular}
\label{tab:coupling}
\end{center}
\end{table*}

In this subsection we construct the $\eta(x,y)$ and $\xi(x,y)$ currents. To do this, we need charmed meson operators as well as their couplings to charmed meson states, which can be found in Table~\ref{tab:coupling} (see Ref.~\cite{Chen:2019wjd} and references therein for detailed discussions). We also need ``ground-state'' charmed baryon fields, which have been systematically constructed and studied in Refs.~\cite{Liu:2007fg,Chen:2017sci,Cui:2019dzj} using the method of QCD sum rules~\cite{Shifman:1978bx,Reinders:1984sr} within the heavy quark effective theory~\cite{Grinstein:1990mj,Eichten:1989zv,Falk:1990yz}. We briefly summarize the results here.

The interpolating fields coupling to the $J^P = 1/2^+$ ground-state charmed baryons $\Lambda_c$ and $\Sigma_c$ are
\begin{eqnarray}
\nonumber J_{\Lambda_c^+} &=& \epsilon^{abc} [u_a^T \mathbb{C} \gamma_{5} d_b] c_c \, ,
\\
\sqrt2 J_{\Sigma_c^{++}} &=& \epsilon^{abc} [u_a^T \mathbb{C} \gamma_{\mu} u_b] \gamma^{\mu}\gamma_{5} c_c \, ,
\label{eq:heavybaryon}
\\ \nonumber
J_{\Sigma_c^{+}}  &=& \epsilon^{abc} [u_a^T \mathbb{C} \gamma_{\mu} d_b] \gamma^{\mu}\gamma_{5} c_c \, ,
\\ \nonumber
\sqrt2 J_{\Sigma_c^{0}}  &=& \epsilon^{abc} [d_a^T \mathbb{C} \gamma_{\mu} d_b] \gamma^{\mu}\gamma_{5} c_c \, .
\end{eqnarray}
Their couplings are defined as
\begin{equation}
\langle 0 | J_{\mathcal{B}} | \mathcal{B} \rangle = f_{\mathcal{B}} u_{\mathcal{B}} \, ,
\end{equation}
where $u_{\mathcal{B}}$ is the Dirac spinor of the charmed baryon ${\mathcal{B}}$, and the decay constants $f_{\mathcal{B}}$ have been calculated in Refs.~\cite{Liu:2007fg,Chen:2017sci,Cui:2019dzj} to be
\begin{eqnarray}
f_{\Lambda_c} &=& 0.015 {\rm~GeV}^3 \, ,
\\
\nonumber f_{\Sigma_c} &=& 0.036 {\rm~GeV}^3 \, .
\end{eqnarray}
The above results are evaluated within the heavy quark effective theory, but for light baryon fields we shall use full QCD decay constants (see Sec.~\ref{sec:current2}). This causes some, but not large, theoretical uncertainties.

Actually, there are several other charmed baryon fields, such as:
\begin{itemize}

\item the ``ground-state'' field of pure $J^P = 3/2^+$
\begin{equation}
J^{\mu}_{\Sigma_c^{*+}} = P_{3/2}^{\mu\alpha} ~ \epsilon^{abc} [u_a^T \mathbb{C} \gamma_\alpha d_b] c_c \, ,
\end{equation}
which couples to the $J^P = 3/2^+$ ground-state charmed baryons $\Sigma_c^{*+}$, with $P_{3/2}^{\mu\alpha}$ the $J=3/2$ projection operator
\begin{equation}
P_{3/2}^{\mu\alpha} = g^{\mu\alpha} - {\gamma^\mu\gamma^\alpha\over4} \, .
\label{eq:projector1}
\end{equation}

\item the ``excited'' charmed baryon field
\begin{equation}
J^*_{\mathcal{B}} = \epsilon^{abc} [u_a^T \mathbb{C} d_b] \gamma_{5} c_c \, ,
\end{equation}
which contains the excited diquark field $\epsilon^{abc} u_a^T \mathbb{C} d_b$ of $J^P = 0^-$.

\end{itemize}
For completeness, we list all of them in Appendix~\ref{app:baryon}, and refer to Ref.~\cite{Dmitrasinovic} for detailed discussions. The major advantage of using the heavy quark effective theory is that within this framework all these charmed baryon fields do not couple to the $J^P = 1/2^+$ ground-state charmed baryons $\Lambda_c$ and $\Sigma_c$~\cite{groundbaryon}. However, some of them, both ``ground-state'' and ``excited'' fields, can couple to the $J^P = 3/2^+$ ground-state charmed baryon $\Sigma_c^{*}$. Hence, we do/can not study decays of $P_c$ states into the $\bar D \Sigma_c^*$ final state in the present study.

Combing charmed meson operators and ground-state charmed baryon fields, we can construct the $\eta(x,y)$ and $\xi(x,y)$ currents. In the molecular picture the $P_c(4312)$, $P_c(4440)$, and $P_c(4457)$ can be interpreted as the $\bar D \Sigma_c$ hadronic molecular state of $J^P = {1/2}^-$, the $\bar D^* \Sigma_c$ one of $J^P = {1/2}^-$, and the $\bar D^* \Sigma_c$ one of $J^P = {3/2}^-$~\cite{Wu:2012md,Chen:2019asm,Liu:2019tjn}:
\begin{eqnarray}
&& | \bar D \Sigma_c; {1/2}^- ; \theta_1 \rangle
\label{def:molecule1}
\\ \nonumber && ~~~~ = \cos\theta_1~| \bar D^0 \Sigma_c^+ \rangle_{J=1/2} + \sin\theta_1~| \bar D^- \Sigma_c^{++} \rangle_{J=1/2} \, ,
\\[1mm] && | \bar D^* \Sigma_c; {1/2}^- ; \theta_2 \rangle
\label{def:molecule2}
\\ \nonumber && ~~~~ = \cos\theta_2~| \bar D^{*0} \Sigma_c^+ \rangle_{J=1/2} + \sin\theta_2~| \bar D^{*-} \Sigma_c^{++} \rangle_{J=1/2} \, ,
\\[1mm] && | \bar D^* \Sigma_c; {3/2}^- ; \theta_3 \rangle
\label{def:molecule3}
\\ \nonumber && ~~~~ = \cos\theta_3~| \bar D^{*0} \Sigma_c^+ \rangle_{J=3/2} + \sin\theta_3~| \bar D^{*-} \Sigma_c^{++} \rangle_{J=3/2} \, ,
\end{eqnarray}
where $\theta_i$ ($i=1,2,3$) are isospin parameters ($\theta_i = -55^{\rm o}$ for $I=1/2$ and $\theta_i = 35^{\rm o}$ for $I=3/2$). Their relevant interpolating currents are:
\begin{eqnarray}
J_i^{(\alpha)} &=& \cos\theta_i~\eta_i^{(\alpha)} + \sin\theta_i~\xi_i^{(\alpha)} \, ,
\end{eqnarray}
where
\begin{eqnarray}
\eta_1 &=& [\bar c_a \gamma_5 u_a] ~ \Sigma_c^+
\\ \nonumber &=& [\delta^{ab} \bar c_a \gamma_5 u_b] ~ [\epsilon^{cde} u_c^T \mathbb{C} \gamma_\mu d_d \gamma^\mu \gamma_5 c_e] \, ,
\\[1mm] \eta_2 &=& [\bar c_a \gamma_\nu u_a] ~ \gamma^\nu \gamma_5 \Sigma_c^+
\\ \nonumber &=& [\delta^{ab} \bar c_a \gamma_\nu u_b] ~ \gamma^\nu \gamma_5 ~ [\epsilon^{cde} u_c^T \mathbb{C} \gamma_\mu d_d \gamma^\mu \gamma_5 c_e] \, ,
\\[1mm] \eta_3^\alpha &=& P_{3/2}^{\alpha\nu} ~ [\bar c_a \gamma_\nu u_a] ~ \Sigma_c^+
\\ \nonumber &=&[\delta^{ab} \bar c_a \gamma_\nu u_b] ~ P_{3/2}^{\alpha\nu}[\epsilon^{cde} u_c^T \mathbb{C} \gamma_\mu d_d \gamma^\mu \gamma_5 c_e] \, ,
\end{eqnarray}
and
\begin{eqnarray}
\xi_1 &=& [\bar c_a \gamma_5 d_a] ~ \Sigma_c^{++}
\\ \nonumber &=& {1\over\sqrt2} ~ [\delta^{ab} \bar c_a \gamma_5 d_b] ~ [\epsilon^{cde} u_c^T \mathbb{C} \gamma_\mu u_d \gamma^\mu \gamma_5 c_e] \, ,
\\[1mm] \xi_2 &=& [\bar c_a \gamma_\nu d_a] ~ \gamma^\nu \gamma_5 \Sigma_c^{++}
\\ \nonumber &=& {1\over\sqrt2} ~ [\delta^{ab} \bar c_a \gamma_\nu d_b] ~ \gamma^\nu \gamma_5 ~ [\epsilon^{cde} u_c^T \mathbb{C} \gamma_\mu u_d \gamma^\mu \gamma_5 c_e] \, ,
\\[1mm] \xi_3^\alpha &=& P_{3/2}^{\alpha\nu} ~ [\bar c_a \gamma_\nu d_a] ~ \Sigma_c^{++}
\\ \nonumber &=& {1\over\sqrt2} ~ [\delta^{ab} \bar c_a \gamma_\nu d_b] ~ P_{3/2}^{\alpha\nu}[\epsilon^{cde} u_c^T \mathbb{C} \gamma_\mu u_d \gamma^\mu \gamma_5 c_e] \, .
\end{eqnarray}
In the above expressions we have written $J_{\mathcal{B}}$ as $\mathcal{B}$ for simplicity.

\subsection{$\theta(x,y)$ and light baryon fields}
\label{sec:current2}

In this subsection we construct the $\theta(x,y)$ currents, which can be constructed by combing charmonium operators and light baryon fields. Hence, we need charmonium operators as well as their couplings to charmonium states, which can be found in Table~\ref{tab:coupling} (see Ref.~\cite{Chen:2019wjd} and references therein for detailed discussions). We also need light baryon fields, which have been systematically studied in Refs.~\cite{Ioffe:1981kw,Ioffe:1982ce,Espriu:1983hu,Chen:2008qv,Chen:2009sf,Chen:2010ba,Chen:2011rh,Dmitrasinovic:2016hup}. We briefly summarize the results here.

According to the results of Ref.~\cite{Chen:2008qv}, we can use $u$, $u$, and $d$ ($q=u/d$) quarks to construct five independent baryon fields:
\begin{eqnarray}
\nonumber N_1 &=& \epsilon^{abc} (u_a^T \mathbb{C} d_b) \gamma_5 u_c \, ,
\\[1mm] \nonumber N_2 &=& \epsilon^{abc} (u_a^T \mathbb{C} \gamma_5 d_b) u_c \, ,
\\[1mm] N_3^\mu &=& P_{3/2}^{\mu\alpha} ~ \epsilon^{abc} (u_a^T \mathbb{C} \gamma_\alpha \gamma_5 d_b) \gamma_5 u_c \, ,
\label{eq:lightbaryon}
\\[1mm] \nonumber N_4^\mu &=& P_{3/2}^{\mu\alpha} ~ \epsilon^{abc} (u_a^T \mathbb{C} \gamma_\alpha d_b) u_c \, ,
\\[1mm] \nonumber N_5^{\mu\nu} &=& P_{3/2}^{\mu\nu\alpha\beta} ~ \epsilon^{abc} (u_a^T \mathbb{C} \sigma_{\alpha\beta} d_b) \gamma_5 u_c \, ,
\end{eqnarray}
where the projection operator $P_{3/2}^{\mu\nu\alpha\beta}$ is
\begin{eqnarray}
\nonumber P_{3/2}^{\mu\nu\alpha\beta} &=& {g^{\mu\alpha}g^{\nu\beta}\over2}  - {g^{\mu\beta}g^{\nu\alpha}\over2}
- {g^{\mu\alpha}\over4}\gamma^\nu\gamma^\beta + {g^{\mu\beta}\over4}\gamma^\nu\gamma^\alpha
\\ && + {g^{\nu\alpha}\over4}\gamma^\mu\gamma^\beta - {g^{\nu\beta}\over4}\gamma^\mu\gamma^\alpha + {1\over6} \sigma^{\mu\nu}\sigma^{\alpha\beta} \, .
\label{eq:projector2}
\end{eqnarray}
All the other light baryon fields \Big(including other $\epsilon^{abc }[u^T_a \mathbb{C} \Gamma_1 d_b]\Gamma_2 u_c$ fields as well as all the $\epsilon^{abc }[u^T_a \mathbb{C} \Gamma_3 u_b]\Gamma_4 d_c$ fields\Big) can be transformed to $N^{(\mu\nu)}_{1,2,3,4,5}$, as shown in Appendix~\ref{app:baryon}.

Among the five fields defined in Eqs.~(\ref{eq:lightbaryon}), the former two $N_{1,2}$ have pure spin $J=1/2$, and the latter three $N^{\mu(\nu)}_{3,4,5}$ have pure spin $J=3/2$. In the present study we shall study decays of $P_c$ states into charmonia and protons, but not study their decays into charmonia and $\Delta/N^*$, since the couplings of $N^{\mu(\nu)}_{3,4,5}$ to $\Delta/N^*$ have not been (well) investigated in the literature. Therefore, we only keep $N_{1,2}$ but omit $N^{\mu(\nu)}_{3,4,5}$. Moreover, we shall find that all the terms in our calculations do not depend on $N_1 + N_2$, so we only need to consider the Ioffe's light baryon field
\begin{equation}
N \equiv N_1 - N_2 \, .
\end{equation}
This field has been well studied in Refs.~\cite{Ioffe:1981kw,Ioffe:1982ce,Espriu:1983hu} and suggested to couple to the proton through
\begin{equation}
\langle 0 | N | p \rangle = f_p u_p \, ,
\end{equation}
with the decay constant evaluated in Ref.~\cite{Chen:2012ex} to be
\begin{equation}
f_p = 0.011  {\rm~GeV}^3 \, .
\end{equation}

\section{Fierz rearrangement}
\label{sec:fierz}

%
\begin{figure*}[hbt]
\begin{center}
\subfigure[~\mbox{$\eta \rightarrow \theta$}]{\includegraphics[width=0.32\textwidth]{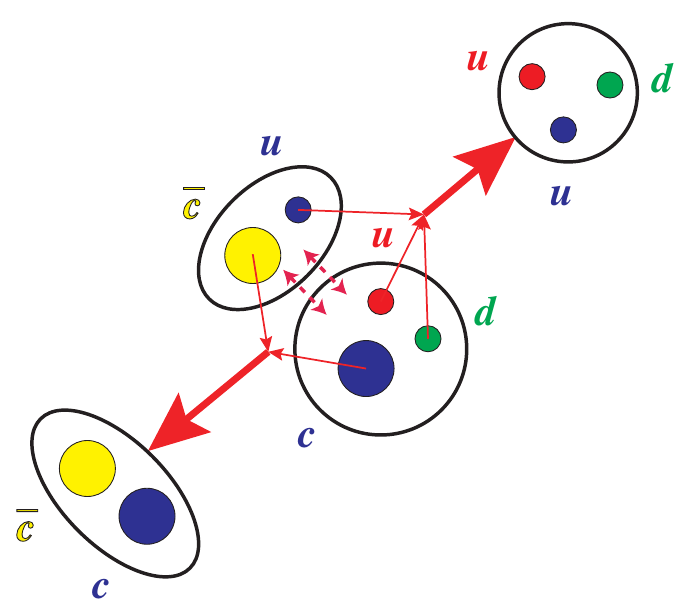}}
\subfigure[~\mbox{$\eta \rightarrow \eta$}]{\includegraphics[width=0.32\textwidth]{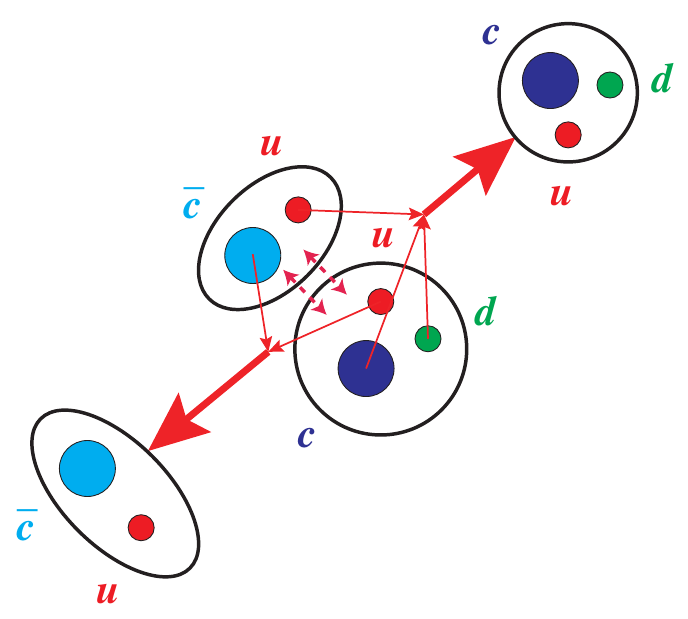}}
\subfigure[~\mbox{$\eta \rightarrow \xi$}]{\includegraphics[width=0.32\textwidth]{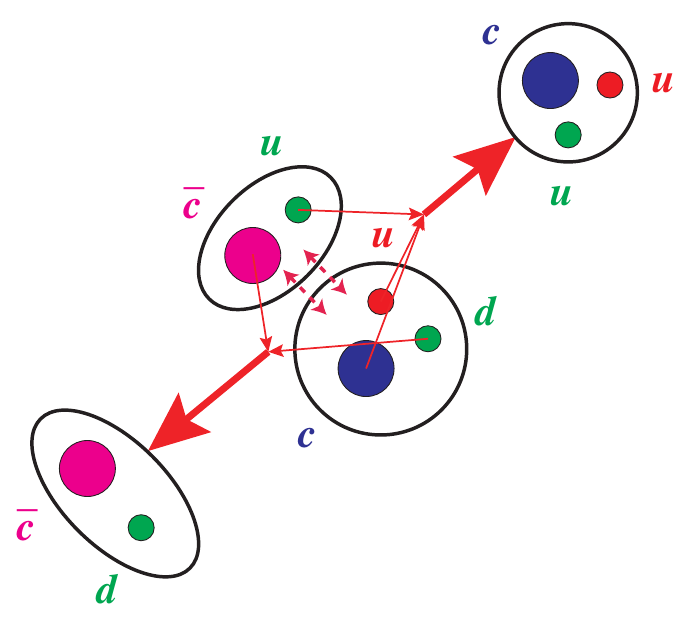}}
\caption{Fall-apart decays of $P_c$ states as $\bar D^{(*)0} \Sigma_c^{+}$ molecular states, investigated through the $\eta(x,y)$ currents. There are three possibilities: a) $\eta \rightarrow \theta$, b) $\eta \rightarrow \eta$, and c) $\eta \rightarrow \xi$. Their probabilities are the same (33\%), if only considering the color degree of freedom.}
\label{fig:decayeta}
\end{center}
\end{figure*}
%

%
\begin{figure*}[hbt]
\begin{center}
\subfigure[~\mbox{$\xi \rightarrow \theta$}]{\includegraphics[width=0.32\textwidth]{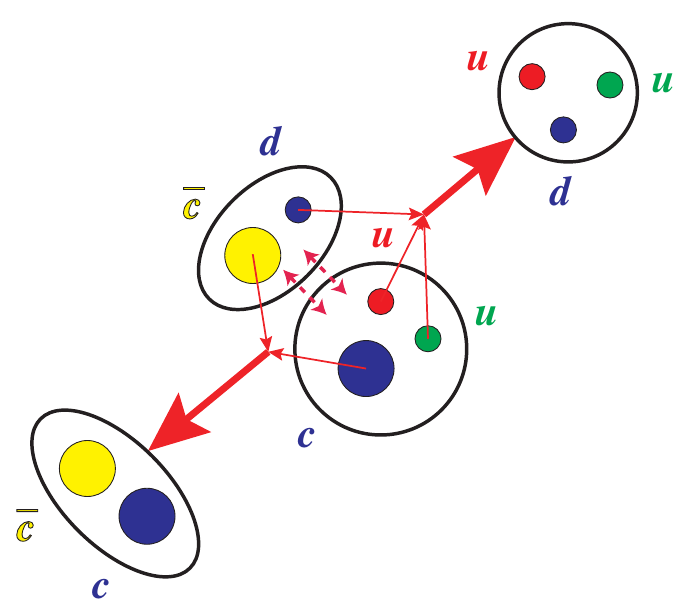}}
\subfigure[~\mbox{$\xi \rightarrow \eta$}]{\includegraphics[width=0.32\textwidth]{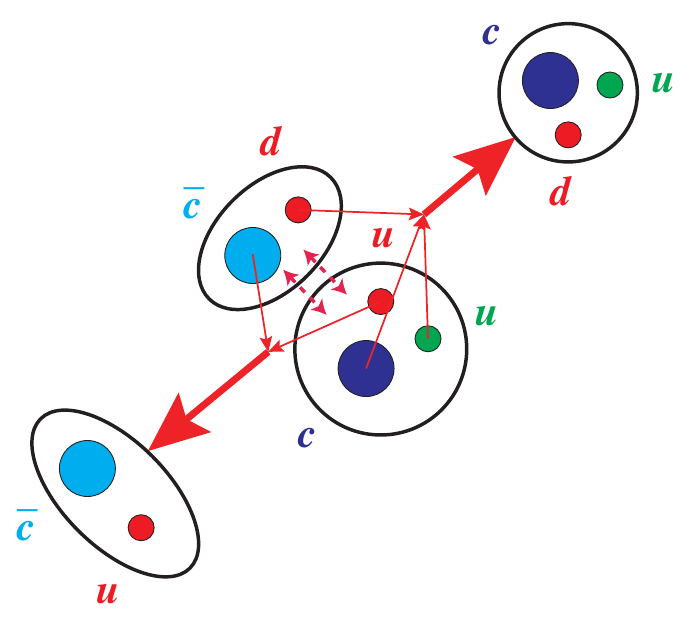}}
\subfigure[~\mbox{$\xi \rightarrow \eta$}]{\includegraphics[width=0.32\textwidth]{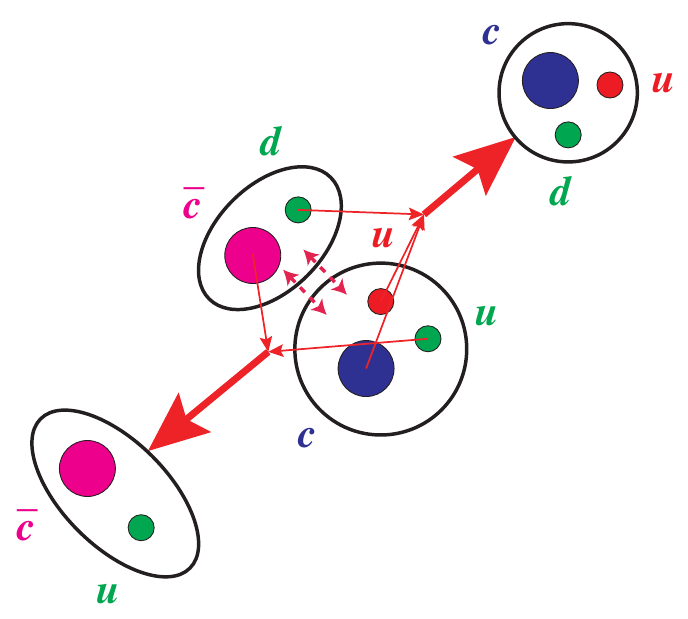}}
\caption{Fall-apart decays of $P_c$ states as $\bar D^{(*)-} \Sigma_c^{++}$ molecular states, investigated through the $\xi(x,y)$ currents. There are three possibilities: a) $\xi \rightarrow \theta$, b) $\xi \rightarrow \eta$, and c) again $\xi \rightarrow \eta$. Their probabilities are the same (33\%), if only considering the color degree of freedom.}
\label{fig:decayxi}
\end{center}
\end{figure*}
%

In this section we study the Fierz rearrangement of the $\eta(x,y)$ and $\xi(x,y)$ currents, which will be used to investigate fall-apart decays of $P_c$ states in Sec.~\ref{sec:decay}. Taking $\eta(x,y)$ as an example, when the $\bar c_a(x)$ and $c_e(y)$ quarks meet each other and the $u_b(x)$, $u_c(y)$, and $d_d(y)$ quarks meet together at the same time, a $\bar D^{(*)0} \Sigma_c^{+}$ molecular state can decay into one charmonium meson and one light baryon. This is the decay process depicted in Fig.~\ref{fig:decayeta}(a):
\begin{eqnarray}
&& \left[\delta^{ab} \bar c_a(x) u_b(x)\right] ~ \left[\epsilon^{cde} u_c(y) d_d(y) c_e(y)\right]
\label{eq:change}
\\ \nonumber &\rightarrow& \delta^{ab} \bar c_a(x \to x^\prime) u_b(x \to y^\prime)
\\ \nonumber && ~~~~~~~~~~~~ \otimes \epsilon^{cde} u_c(y \to y^\prime) d_d(y \to y^\prime) c_e(y \to x^\prime)
\\ \nonumber &=& {1\over3}\delta^{ae} \epsilon^{bcd} \otimes \bar c_a(x^\prime)u_b(y^\prime) \otimes u_c(y^\prime) d_d(y^\prime) c_e(x^\prime) + \cdots
\\ \nonumber &=& {1\over3} \left[\delta^{ae}\bar c_a(x^\prime) c_e(x^\prime)\right] \oplus \left[\epsilon^{bcd}u_b(y^\prime) u_c(y^\prime) d_d(y^\prime)\right] + \cdots .
\end{eqnarray}
The first step is a dynamical process, during which we assume that all the color, flavor, spin and orbital structures remain unchanged, so the relevant current also remains the same. The second and third steps can be described by applying the Fierz rearrangement to interchange both the color and Dirac indices of the $u_b(y^\prime)$ and $c_e(x^\prime)$ quark fields.

Still taking $\eta(x,y)$ as an example: when the $\bar c_a(x)$ and $u_c(y)$ quarks meet each other and the $u_b(x)$, $d_d(y)$, and $c_e(y)$ quarks meet together at the same time, a $\bar D^{(*)0} \Sigma_c^{+}$ molecular state can decay into one charmed meson and one charmed baryon, as depicted in Fig.~\ref{fig:decayeta}(b); when the $\bar c_a(x)$ and $d_d(y)$ quarks meet each other and the $u_b(x)$, $u_c(y)$, and $c_e(y)$ quarks meet together at the same time, a $\bar D^{(*)0} \Sigma_c^{+}$ molecular state can also decay into one charmed meson and one charmed baryon, as depicted in Fig.~\ref{fig:decayeta}(c). Similarly, decays of $\bar D^{(*)-} \Sigma_c^{++}$ molecular states can be investigated through the $\xi(x,y)$ currents, as depicted in Fig.~\ref{fig:decayxi}(a,b,c).

In the following subsections we shall study the above fall-apart decay processes, by applying the Fierz rearrangement~\cite{fierz} of the Dirac and color indices to relate the $\eta$, $\xi$, and $\theta$ currents. This method has been used to systematically study light baryon and tetraquark operators/currents in Refs.~\cite{Chen:2008qv,Chen:2009sf,Chen:2010ba,Chen:2011rh,Dmitrasinovic:2016hup,Chen:2006hy,Chen:2006zh,Chen:2007xr,Chen:2008ej,Chen:2008qw,Chen:2013jra,Chen:2018kuu}. We note that the Fierz rearrangement in the Lorentz space is actually a matrix identity, which is valid if each quark field in the initial and final operators is at the same location, {\it e.g.}, we can apply the Fierz rearrangement to transform a non-local $\eta$ current with the quark fields $\eta = [\bar c(x^\prime) u(y^\prime)] ~ [u(y^\prime) d(y^\prime) c(x^\prime)]$ into the combination of many non-local $\theta$ currents with the quark fields at same locations $\theta = [\bar c(x^\prime) c(x^\prime)] ~ [u(y^\prime) u(y^\prime) d(y^\prime)]$. Hence, this rearrangement exactly describes the third step of Eq.~(\ref{eq:change}).

\subsection{$\eta \rightarrow \theta$ and $\xi \rightarrow \theta$}

Using Eq.~(\ref{eq:color}), together with the Fierz rearrangement to interchange the $u_b$ and $c_e$ quark fields, we can transform an $\eta(x,y)$ current into the combination of many $\theta$ currents:
\begin{eqnarray}
\eta_1 &\rightarrow& {1\over12} ~ [\bar c_a c_a]~\gamma_5 N - {1\over12} ~ [\bar c_a \gamma_5 c_a]~N
\label{eq:eta1theta}
\\ \nonumber &&  + {1\over24} ~ [\bar c_a \gamma_\mu \gamma_5 c_a]~\gamma^\mu N  + {1\over24} ~ [\bar c_a \gamma_\mu c_a]~\gamma^\mu \gamma_5 N
\\ \nonumber &&  +~\cdots \, ,
\\
\eta_2 &\rightarrow& {1\over6} ~ [\bar c_a c_a]~\gamma_5 N + {1\over6} ~ [\bar c_a \gamma_5 c_a]~N
\label{eq:eta2theta}
\\ \nonumber &&  + {1\over12} ~ [\bar c_a \gamma_\mu \gamma_5 c_a]~\gamma^\mu N - {1\over12} ~ [\bar c_a \gamma_\mu c_a]~\gamma^\mu \gamma_5 N
\\ \nonumber &&  - {1\over12} ~ [\bar c_a \sigma_{\mu\nu} c_a]~\sigma^{\mu\nu} \gamma_5 N~+~\cdots \, ,
\\
\eta_3^\alpha &\rightarrow& [\bar c_a \gamma_\mu \gamma_5 c_a] ~ \left({1\over16} g^{\alpha\mu} \gamma_5 + {i\over48} \sigma^{\alpha\mu} \gamma_5 \right) N
\label{eq:eta3theta}
\\ \nonumber &&  + ~ [\bar c_a \gamma_\mu c_a] ~ \left(-{1\over16} g^{\alpha\mu} - {i\over48} \sigma^{\alpha\mu} \right) N~+~\cdots  \, .
\end{eqnarray}
In the above transformations we have changed the coordinates according to the first step of Eq.~(\ref{eq:change}), which are not shown explicitly here for simplicity. Besides, we have omitted in $\cdots$ that: a) the color-octet-color-octet meson-baryon terms, and b) terms depending on the $J=3/2$ light baryon fields $N^{\mu(\nu)}_{3,4,5}$. Hence, we have only kept, but kept all, the color-singlet-color-singlet meson-baryon terms depending on the $J=1/2$ fields $N_1$ and $N_2$. {\it This is not an easy task because we need to use many identities given in Eqs.~(\ref{eq:lightbaryon1}) and (\ref{eq:lightbaryon2}) of Appendix~\ref{app:baryon} in order to safely omit $N^{\mu(\nu)}_{3,4,5}$}. Moreover, we can find in the above expressions that all terms contain the Ioffe's light baryon field $N \equiv N_1 - N_2$, and there are no terms depending on $N_1 + N_2$.

The above transformations can be used to describe the fall-apart decay process depicted in Fig.~\ref{fig:decayeta}(a) for $\bar D^{(*)0} \Sigma_c^{+}$ molecular states. Similarly, we can investigate the fall-apart decay process depicted in Fig.~\ref{fig:decayxi}(a) for $\bar D^{(*)-} \Sigma_c^{++}$ molecular states. To do this, we need to use Eq.~(\ref{eq:color}), together with the Fierz rearrangement to interchange the $d_b$ and $c_e$ quark fields, to transform a $\xi(x,y)$ current into the combination of many $\theta$ currents:
\begin{eqnarray}
\sqrt2\xi_1 &\rightarrow& - {1\over6} ~ [\bar c_a c_a]~\gamma_5 N + {1\over6} ~ [\bar c_a \gamma_5 c_a]~N
\label{eq:xi1theta}
\\ \nonumber && - {1\over12} ~ [\bar c_a \gamma_\mu \gamma_5 c_a]~\gamma^\mu N - {1\over12} ~ [\bar c_a \gamma_\mu c_a]~\gamma^\mu \gamma_5 N
\\ \nonumber && +~\cdots \, ,
\\
\sqrt2\xi_2 &\rightarrow& - {1\over3} ~ [\bar c_a c_a]~\gamma_5 N  - {1\over3} ~ [\bar c_a \gamma_5 c_a]~N
\label{eq:xi2theta}
\\ \nonumber &&  - {1\over6} ~ [\bar c_a \gamma_\mu \gamma_5 c_a]~\gamma^\mu N + {1\over6} ~ [\bar c_a \gamma_\mu c_a]~\gamma^\mu \gamma_5 N
\\ \nonumber &&  + {1\over6} ~ [\bar c_a \sigma_{\mu\nu} c_a] ~ \sigma^{\mu\nu} \gamma_5 N  ~+~\cdots  \, ,
\\
\sqrt2\xi_3^\alpha &\rightarrow&  [\bar c_a \gamma_\mu \gamma_5 c_a] ~ \left(-{1\over8} g^{\alpha\mu} \gamma_5 - {i\over24} \sigma^{\alpha\mu} \gamma_5 \right)N
\label{eq:xi3theta}
\\ \nonumber &&  + ~ [\bar c_a \gamma_\mu c_a] ~ \left({1\over8} g^{\alpha\mu} + {i\over24} \sigma^{\alpha\mu} \right)N  ~+~\cdots  \, .
\end{eqnarray}

\subsection{$\eta \rightarrow \eta$ and $\eta \rightarrow \xi$}

First we derive a color rearrangement similar to Eq.~(\ref{eq:color}):
\begin{equation}
\delta^{ab} \epsilon^{cde} = {1\over3}~\delta^{ac} \epsilon^{bde} - {1\over2}~\lambda^{ac}_n \epsilon^{bdf} \lambda^{fe}_n + {1\over2}~\lambda^{ac}_n \epsilon^{bef} \lambda^{fd}_n \, .
\label{eq:color2}
\end{equation}
Using this identity, together with the Fierz rearrangement to interchange the $u_b$ and $u_c$ quark fields, we can transform an $\eta(x,y)$ current into the combination of many $\eta$ currents.

Besides, we can derive another similar color rearrangement:
\begin{equation}
\delta^{ab} \epsilon^{cde} = {1\over3}~\delta^{ad} \epsilon^{cbe} + {1\over2}~\lambda^{ad}_n \epsilon^{bcf} \lambda^{fe}_n - {1\over2}~\lambda^{ad}_n \epsilon^{bef} \lambda^{fc}_n \, .
\label{eq:color3}
\end{equation}
Using this identity, together with the Fierz rearrangement to interchange the $u_b$ and $d_d$ quark fields, we can transform an $\eta(x,y)$ current into the combination of many $\xi$ currents.

The above two transformations describe the fall-apart decay processes depicted in Fig.~\ref{fig:decayeta}(b,c) for $\bar D^{(*)0} \Sigma_c^{+}$ molecular states. Altogether, we obtain:
\begin{eqnarray}
\eta_1 &\rightarrow& {1\over12}~[\bar c_a \gamma_\mu u_a] ~ \gamma^\mu \gamma_5 \Lambda_c^+
\label{eq:eta1etaxi}
\\ \nonumber &&    - {1\over12}~[\bar c_a \gamma_5 u_a] ~ \Sigma_c^{+} - {\sqrt2\over12}~[\bar c_a \gamma_5 d_a] ~ \Sigma_c^{++}
\\ \nonumber &&  - {1\over24}~[\bar c_a \sigma_{\mu\nu} u_a] ~ \epsilon^{\mu\nu\rho\sigma} \gamma_\sigma \gamma_5 \left( - {1\over4} \gamma_\rho \gamma_5 \Sigma_c^{+} \right)
\\ \nonumber &&  - {\sqrt2\over24}~[\bar c_a \sigma_{\mu\nu} d_a] \epsilon^{\mu\nu\rho\sigma} \gamma_\sigma \gamma_5 \left( - {1\over4} \gamma_\rho \gamma_5 \Sigma_c^{++} \right) + \cdots \, ,
\\
\eta_2 &\rightarrow& {1\over3}~[\bar c_a \gamma_5 u_a] ~ \Lambda_c^+ - {1\over12}~[\bar c_a \sigma_{\mu\nu} u_a] ~ \sigma^{\mu\nu} \gamma_5 \Lambda_c^+
\label{eq:eta2etaxi}
\\ \nonumber &&   - {1\over6} [\bar c_a \gamma_\mu u_a] ~ \left( - {1\over4} \gamma^\mu \gamma_5 \Sigma_c^{+} \right)
\\ \nonumber &&   - {i\over6} [\bar c_a \gamma_\mu \gamma_5 u_a] ~ \sigma^{\mu\nu} \gamma_5  \left( - {1\over4} \gamma_\nu \gamma_5 \Sigma_c^{+} \right)
\\ \nonumber &&   - {\sqrt2\over6} [\bar c_a \gamma_\mu d_a] ~ \left( - {1\over4} \gamma^\mu \gamma_5 \Sigma_c^{++} \right)
\\ \nonumber &&   - {i\sqrt2\over6} [\bar c_a \gamma_\mu \gamma_5 d_a] ~ \sigma^{\mu\nu} \gamma_5  \left( - {1\over4} \gamma_\nu \gamma_5 \Sigma_c^{++} \right) + \cdots \, ,
\\
\eta_3^\alpha &\rightarrow&  \left(- {i\over48} g^{\alpha\mu} \gamma^\nu + {i\over48} g^{\alpha\nu} \gamma^\mu - {1\over48} \epsilon^{\alpha\beta\mu\nu} \gamma_\beta \gamma_5 \right)
\label{eq:eta3etaxi}
\\ \nonumber && ~~~\times~[\bar c_a \sigma_{\mu\nu} u_a] ~ \Lambda_c^+
\\ \nonumber &&  +~\left(- {1\over12} g^{\alpha\mu} \gamma^\nu \gamma_5 - {1\over12} g^{\alpha\nu} \gamma^\mu \gamma_5 + {1\over24} g^{\mu\nu} \gamma^\alpha \gamma_5 \right)
\\ \nonumber && ~~~\times~[\bar c_a \gamma_\mu u_a] ~  \left( - {1\over4} \gamma_\nu \gamma_5 \Sigma_c^{+} \right)
\\ \nonumber &&  +~\left({1\over24} g^{\alpha\mu} \gamma^\nu - {1\over24} g^{\alpha\nu} \gamma^\mu - {i\over24} \epsilon^{\alpha\beta\mu\nu} \gamma_\beta \gamma_5 \right)
\\ \nonumber && ~~~\times~[\bar c_a \gamma_\mu \gamma_5 u_a] ~  \left( - {1\over4} \gamma_\nu \gamma_5 \Sigma_c^{+} \right)
\\ \nonumber &&  +~\left(- {1\over12} g^{\alpha\mu} \gamma^\nu \gamma_5 - {1\over12} g^{\alpha\nu} \gamma^\mu \gamma_5 + {1\over24} g^{\mu\nu} \gamma^\alpha \gamma_5 \right)
\\ \nonumber && ~~~\times~\sqrt2~[\bar c_a \gamma_\mu d_a] ~  \left( - {1\over4} \gamma_\nu \gamma_5 \Sigma_c^{++} \right)
\\ \nonumber &&  +~\left({1\over24} g^{\alpha\mu} \gamma^\nu - {1\over24} g^{\alpha\nu} \gamma^\mu - {i\over24} \epsilon^{\alpha\beta\mu\nu} \gamma_\beta \gamma_5 \right)
\\ \nonumber && ~~~\times\sqrt2[\bar c_a \gamma_\mu \gamma_5 d_a] \left( - {1\over4} \gamma_\nu \gamma_5 \Sigma_c^{++} \right) +\cdots .
\end{eqnarray}
In the above transformations we have only kept, but kept all, the color-singlet-color-singlet meson-baryon terms depending on the $J^P=1/2^+$ ``ground-state'' charmed baryon fields given in Eqs.~(\ref{eq:heavybaryon}). {\it Again, this is not an easy task because we need to carefully omit the terms depending on the other charmed baryon fields, $B^G_{\mathbf{\bar 3},1}$, $B^G_{\mathbf{\bar 3},3}$, $B^G_{\mathbf{\bar 3},\mu}$, $B^U_{\mathbf{6},5}$, $B^{U}_{\mathbf{6},\mu}$, $B^{\prime U}_{\mathbf{6},\mu}$, and $B^U_{\mathbf{6},\mu\nu}$, whose definitions can be found in Appendix~\ref{app:baryon}}.

\subsection{$\xi \rightarrow \eta$}

Following the procedures used in the previous subsection, we can transform a $\xi(x,y)$ current into the combination of many $\eta$ currents (without $\xi$ currents):
\begin{eqnarray}
\sqrt2\xi_1 &\rightarrow& - {1\over6}~[\bar c_a \gamma_\mu u_a]~\gamma^\mu \gamma_5 \Lambda_c^+ - {1\over6}~[\bar c_a \gamma_5 u_a]~\Sigma_c^{+}
\label{eq:xi1eta}
\\ \nonumber &&  - {1\over12}~[\bar c_a \sigma_{\mu\nu} u_a] ~ \epsilon^{\mu\nu\rho\sigma} \gamma_\sigma \gamma_5 \left( - {1\over4} \gamma_\rho \gamma_5 \Sigma_c^{+} \right)
\\ \nonumber &&  +~\cdots \, ,
\\
\sqrt2\xi_2 &\rightarrow& - {2\over3} ~ [\bar c_a \gamma_5 u_a] ~ \Lambda_c^+ + {1\over6}~[\bar c_a \sigma_{\mu\nu} u_a] ~ \sigma^{\mu\nu} \gamma_5 \Lambda_c^+
\label{eq:xi2eta}
\\ \nonumber &&  - {1\over3}~[\bar c_a \gamma_\mu u_a] ~  \left( - {1\over4} \gamma^\mu \gamma_5 \Sigma_c^{+} \right)
\\ \nonumber &&  - {i\over3}~[\bar c_a \gamma_\mu \gamma_5 u_a] ~ \sigma^{\mu\nu} \gamma_5  \left( - {1\over4} \gamma_\nu \gamma_5 \Sigma_c^{+} \right) + \cdots \, ,
\\
\sqrt2\xi_3^\alpha &\rightarrow&   \left({i\over24} g^{\alpha\mu} \gamma^\nu - {i\over24} g^{\alpha\nu} \gamma^\mu + {1\over24} \epsilon^{\alpha\beta\mu\nu} \gamma_\beta \gamma_5 \right)
\label{eq:xi3eta}
\\ \nonumber && ~~~\times~[\bar c_a \sigma_{\mu\nu} u_a] ~ \Lambda_c^+
\\ \nonumber &&  +~\left(- {1\over6} g^{\alpha\mu} \gamma^\nu \gamma_5 - {1\over6} g^{\alpha\nu} \gamma^\mu \gamma_5 + {1\over12} g^{\mu\nu} \gamma^\alpha \gamma_5 \right)
\\ \nonumber && ~~~\times~[\bar c_a \gamma_\mu u_a] ~  \left( - {1\over4} \gamma_\nu \gamma_5 \Sigma_c^{+} \right)
\\ \nonumber &&  +~\left({1\over12} g^{\alpha\mu} \gamma^\nu - {1\over12} g^{\alpha\nu} \gamma^\mu - {i\over12} \epsilon^{\alpha\beta\mu\nu} \gamma_\beta \gamma_5 \right)
\\ \nonumber && ~~~\times~[\bar c_a \gamma_\mu \gamma_5 u_a] ~  \left( - {1\over4} \gamma_\nu \gamma_5 \Sigma_c^{+} \right)
 +\cdots  \, .
\end{eqnarray}
The above transformations describe the fall-apart decay processes depicted in Fig.~\ref{fig:decayxi}(b,c) for $\bar D^{(*)-} \Sigma_c^{++}$ molecular states.

\section{Decay properties of $\bar D^{(*)0} \Sigma_c^+$ and $\bar D^{(*)-} \Sigma_c^{++}$ molecular states}
\label{sec:decay}

In this section we use the Fierz rearrangements derived in the previous section to extract some strong decay properties of $\bar D^{(*)0} \Sigma_c^+$ and $\bar D^{(*)-} \Sigma_c^{++}$ molecular states. We shall separately investigate:
\begin{itemize}

\item $|\bar D^0 \Sigma_c^{+}; 1/2^- \rangle$, the $\bar D^0 \Sigma_c^{+}$ molecular state of $J^P = 1/2^-$, through the $\eta_1(x,y)$ current and the Fierz rearrangements given in Eqs.~(\ref{eq:eta1theta}) and (\ref{eq:eta1etaxi});

\item $|\bar D^- \Sigma_c^{++}; 1/2^- \rangle$, the $\bar D^- \Sigma_c^{++}$ molecular state of $J^P = 1/2^-$, through the $\xi_1(x,y)$ current and the Fierz rearrangements given in Eqs.~(\ref{eq:xi1theta}) and (\ref{eq:xi1eta});

\item $|\bar D^{*0} \Sigma_c^{+}; 1/2^- \rangle$, the $\bar D^{*0} \Sigma_c^{+}$ molecular state of $J^P = 1/2^-$, through the $\eta_2(x,y)$ current and the Fierz rearrangements given in Eqs.~(\ref{eq:eta2theta}) and (\ref{eq:eta2etaxi});

\item $|\bar D^{*-} \Sigma_c^{++}; 1/2^- \rangle$, the $\bar D^{*-} \Sigma_c^{++}$ molecular state of $J^P = 1/2^-$, through the $\xi_2(x,y)$ current and the Fierz rearrangements given in Eqs.~(\ref{eq:xi2theta}) and (\ref{eq:xi2eta});

\item $|\bar D^{*0} \Sigma_c^{+}; 3/2^- \rangle$, the $\bar D^{*0} \Sigma_c^{+}$ molecular state of $J^P = 3/2^-$, through the $\eta_3^\alpha(x,y)$ current and the Fierz rearrangements given in Eqs.~(\ref{eq:eta3theta}) and (\ref{eq:eta3etaxi});

\item $|\bar D^{*-} \Sigma_c^{++}; 3/2^- \rangle$, the $\bar D^{*-} \Sigma_c^{++}$ molecular state of $J^P = 3/2^-$, through the $\xi_3^\alpha(x,y)$ current and the Fierz rearrangements given in Eqs.~(\ref{eq:xi3theta}) and (\ref{eq:xi3eta}).

\end{itemize}
The obtained results will be combined in Sec.~\ref{sec:isospin} to further study decay properties of $\bar D^{(*)} \Sigma_c$ molecular states with definite isospins.

\subsection{$\eta_1 \rightarrow \theta / \eta / \xi$}
\label{sec:decayeta1}

In this subsection we study strong decay properties of $|\bar D^0 \Sigma_c^{+}; 1/2^- \rangle$ through the $\eta_1(x,y)$ current. First we use the Fierz rearrangement given in Eq.~(\ref{eq:eta1theta}) to study the decay process depicted in Fig.~\ref{fig:decayeta}(a), {\it i.e.}, decays of $|\bar D^0 \Sigma_c^{+}; 1/2^- \rangle$ into one charmonium meson and one light baryon. Together with Table~\ref{tab:coupling}, we extract the following decay channels that are kinematically allowed:
\begin{enumerate}

\item The decay of $|\bar D^0 \Sigma_c^{+}; 1/2^- \rangle$ into $\eta_c p$ is contributed by both $[\bar c_a \gamma_5 c_a]~N$ and $[\bar c_a \gamma_\mu \gamma_5 c_a]~\gamma^\mu N$:
\begin{eqnarray}
&& \langle \bar D^0 \Sigma_c^{+}; 1/2^-(q) ~|~\eta_c(q_1)~p(q_2) \rangle
\\ \nonumber &\approx& {ia_1\over12}~\lambda_{\eta_c} f_p ~ \bar u u_p + {ia_1\over24}~ f_{\eta_c} f_p ~ q_1^\mu \bar u \gamma_\mu u_p
\\ \nonumber &\equiv& A_{\eta_c p}~\bar u u_p + A^\prime_{\eta_c p}~q_1^\mu \bar u \gamma_\mu u_p \, ,
\end{eqnarray}
where $u$ and $u_p$ are the Dirac spinors of the $P_c$ state with $J^P = 1/2^-$ and the proton, respectively; $a_1$ is an overall factor, related to the coupling of $\eta_1(x,y)$ to $|\bar D^0 \Sigma_c^{+}; 1/2^- \rangle$ as well as the dynamical process of Fig.~\ref{fig:decayeta}(a); the two coupling constants $A_{\eta_c p}$ and $A^\prime_{\eta_c p}$ are defined for the two different effective Lagrangians
\begin{eqnarray}
\mathcal{L}_{\eta_c p} &=& A_{\eta_c p}~\bar P_c N~\eta_c \, ,
\label{lag:etacpA}
\\ \mathcal{L}^\prime_{\eta_c p} &=& A^\prime_{\eta_c p}~\bar P_c \gamma_\mu N~\partial^\mu\eta_c \, .
\label{lag:etacpB}
\end{eqnarray}

\item The decay of $|\bar D^0 \Sigma_c^{+}; 1/2^- \rangle$ into $J/\psi p$ is contributed by $[\bar c_a \gamma_\mu c_a]~\gamma^\mu \gamma_5 N$:
\begin{eqnarray}
&& \langle \bar D^0 \Sigma_c^{+}; 1/2^-(q) | J/\psi(q_1,\epsilon_1)~p(q_2) \rangle
\\ \nonumber &\approx& {a_1\over24}~ m_{J/\psi} f_{J/\psi} f_p ~ \epsilon_1^\mu \bar u \gamma_\mu \gamma_5 u_p
\\ \nonumber &\equiv& A_{\psi p}~\epsilon_1^\mu \bar u \gamma_\mu \gamma_5 u_p \, ,
\end{eqnarray}
where $A_{\psi p}$ is defined for
\begin{eqnarray}
\mathcal{L}_{\psi p} &=& A_{\psi p}~\bar P_c \gamma_\mu \gamma_5 N~\psi^\mu \, .
\end{eqnarray}

\end{enumerate}
Then we use the Fierz rearrangement given in Eq.~(\ref{eq:eta1etaxi}) to study the decay processes depicted in Fig.~\ref{fig:decayeta}(b,c), {\it i.e.}, decays of $|\bar D^0 \Sigma_c^{+}; 1/2^- \rangle$ into one charmed meson and one charmed baryon. Together with Table~\ref{tab:coupling}, we extract only one decay channel that is kinematically allowed:
\begin{enumerate}

\item[3.] The decay of $|\bar D^0 \Sigma_c^{+}; 1/2^- \rangle$ into $\bar D^{*0} \Lambda_c^+$ is contributed by $[\bar c_a \gamma_\mu u_a]~\gamma^\mu \gamma_5 \Lambda_c^+$:
\begin{eqnarray}
&& \langle \bar D^0 \Sigma_c^{+}; 1/2^-(q) ~|~\bar D^{*0}(q_1,\epsilon_1)~\Lambda_c^+(q_2) \rangle
\\ \nonumber &\approx& {a_2\over12}~ m_{D^*} f_{D^*} f_{\Lambda_c} ~ \epsilon_1^\mu \bar u \gamma_\mu \gamma_5 u_{\Lambda_c}
\\ \nonumber &\equiv& A_{\bar D^{*} \Lambda_c}~\epsilon_1^\mu \bar u \gamma_\mu \gamma_5 u_{\Lambda_c^+} \, ,
\end{eqnarray}
where $u_{\Lambda_c}$ is the Dirac spinor of the $\Lambda_c^+$; $a_2$ is an overall factor, related to the coupling of $\eta_1(x,y)$ to $|\bar D^0 \Sigma_c^{+}; 1/2^- \rangle$ as well as the dynamical processes of Fig.~\ref{fig:decayeta}(b,c); the coupling constant $A_{\bar D^{*} \Lambda_c}$ is defined for
\begin{eqnarray}
\mathcal{L}_{\bar D^{*} \Lambda_c} &=& A_{\bar D^{*} \Lambda_c}~\bar P_c \gamma_\mu \gamma_5 \Lambda_c^+~\bar D^{*,\mu} \, .
\end{eqnarray}

\end{enumerate}

In the molecular picture the $P_c(4312)$ is usually interpreted as the $\bar D \Sigma_c$ hadronic molecular state of $J^P = {1/2}^-$. Accordingly, we assume the mass of $|\bar D^0 \Sigma_c^{+}; 1/2^- \rangle$ to be 4311.9~MeV (more parameters can be found in Appendix~\ref{app:parameters}), and summarize the above decay amplitudes to obtain the following (relative) decay widths:
\begin{eqnarray}
\nonumber \Gamma(|\bar D^0 \Sigma_c^{+}; 1/2^- \rangle \to \eta_c p ) &=& a_1^2 ~ 1.1 \times 10^{5}~{\rm GeV}^7 \, ,
\\[1mm] \nonumber \Gamma(|\bar D^0 \Sigma_c^{+}; 1/2^- \rangle \to J/\psi p ) &=& a_1^2 ~ 2.8 \times 10^{4}~{\rm GeV}^7 \, ,
\\[1mm] \nonumber \Gamma(|\bar D^0 \Sigma_c^{+}; 1/2^- \rangle \to \bar D^{*0} \Lambda_c^+ ) &=& a_2^2 ~ 2.0 \times 10^{4}~{\rm GeV}^7 \, .
\\ \label{result:eta1}
\end{eqnarray}

There are two different effective Lagrangians for the $|\bar D^0 \Sigma_c^{+}; 1/2^- \rangle$ decays into the $\eta_c p$ final state, as given in Eqs.~(\ref{lag:etacpA}) and (\ref{lag:etacpB}). It is interesting to see their individual contributions:
\begin{eqnarray}
\nonumber \Gamma(|\bar D^0 \Sigma_c^{+}; 1/2^- \rangle \to \eta_c p )\big|_{\mathcal{L}_{\eta_c p}} &=& a_1^2 ~ 4.9 \times 10^{4}~{\rm GeV}^7 \, ,
\\[1mm]
\nonumber \Gamma(|\bar D^0 \Sigma_c^{+}; 1/2^- \rangle \to \eta_c p )\big|_{\mathcal{L}^\prime_{\eta_c p}} &=& a_1^2 ~ 1.1 \times 10^{4}~{\rm GeV}^7 \, .
\\
\end{eqnarray}
Hence, the former is about four times larger than the latter. We note that their interference can be important, but the phase angle between them, {\it i.e.}, the phase angle between the two coupling constants $A_{\eta_c p}$ and $A^\prime_{\eta_c p}$, can not be well determined in the present study. We shall investigate its relevant uncertainty in Appendix~\ref{app:phase}.

\subsection{$\xi_1 \rightarrow \theta / \eta$}
\label{sec:decayxi1}

In this subsection we follow the procedures used in the previous subsection to study decay properties of $|\bar D^{-} \Sigma_c^{++}; 1/2^- \rangle$, through the $\xi_1(x,y)$ current and the Fierz rearrangements given in Eqs.~(\ref{eq:xi1theta}) and (\ref{eq:xi1eta}). Again, we assume its mass to be 4311.9~MeV, and obtain the following (relative) decay widths:
\begin{eqnarray}
\nonumber \Gamma(|\bar D^- \Sigma_c^{++}; 1/2^- \rangle \to \eta_c p ) &=& b_1^2 ~ 2.1 \times 10^{5}~{\rm GeV}^7 \, ,
\\[1mm] \nonumber \Gamma(|\bar D^- \Sigma_c^{++}; 1/2^- \rangle \to J/\psi p ) &=& b_1^2 ~ 5.7 \times 10^{4}~{\rm GeV}^7 \, ,
\\[1mm] \nonumber \Gamma(|\bar D^- \Sigma_c^{++}; 1/2^- \rangle \to \bar D^{*0} \Lambda_c^+ ) &=& b_2^2 ~ 3.9 \times 10^{4}~{\rm GeV}^7 \, .
\\ \label{result:xi1}
\end{eqnarray}
Here $b_1$ and $b_2$ are two overall factors, which we simply assume to be $b_1 = a_1$ and $b_2 = a_2$ in the following analyses.

The above widths of the $|\bar D^{-} \Sigma_c^{++}; 1/2^- \rangle$ decays into the $\eta_c p$, $J/\psi p$, and $\bar D^{*0} \Lambda_c^+$ final states are all two times larger than those given in Eqs.~(\ref{result:eta1}) for the $|\bar D^{0} \Sigma_c^{+}; 1/2^- \rangle$ decays.

\subsection{$\eta_2 \rightarrow \theta / \eta / \xi$}
\label{sec:decayeta2}

In this subsection we follow the procedures used in Sec.~\ref{sec:decayeta1} to study decay properties of $|\bar D^{*0} \Sigma_c^{+}; 1/2^- \rangle$ through the $\eta_2(x,y)$ current. First we use the Fierz rearrangement given in Eq.~(\ref{eq:eta2theta}) to study the decay process depicted in Fig.~\ref{fig:decayeta}(a):
\begin{enumerate}

\item The decay of $|\bar D^{*0} \Sigma_c^{+}; 1/2^- \rangle$ into $\eta_c p$ is
\begin{eqnarray}
&& \langle \bar D^{*0} \Sigma_c^{+}; 1/2^-(q) ~|~\eta_c(q_1)~p(q_2) \rangle
\\ \nonumber &\approx& - {ic_1\over6}~\lambda_{\eta_c} f_p ~ \bar u u_p + {ic_1\over12}~ f_{\eta_c} f_p ~ q_1^\mu \bar u \gamma_\mu u_p
\\ \nonumber &\equiv& C_{\eta_c p}~\bar u u_p + C^\prime_{\eta_c p}~q_1^\mu \bar u \gamma_\mu u_p \, ,
\end{eqnarray}
where $c_1$ is an overall factor.

\item The decay of $|\bar D^{*0} \Sigma_c^{+}; 1/2^- \rangle$ into $J/\psi p$ is contributed by both $[\bar c_a \gamma_\mu c_a]~\gamma^\mu \gamma_5 N$ and $[\bar c_a \sigma_{\mu\nu} c_a]~\sigma^{\mu\nu} \gamma_5 N$:
\begin{eqnarray}
&& \langle \bar D^{*0} \Sigma_c^{+}; 1/2^-(q) | J/\psi(q_1,\epsilon_1)~p(q_2) \rangle
\\ \nonumber &\approx& - {c_1\over12}~ m_{J/\psi} f_{J/\psi} f_p ~ \epsilon_1^\mu \bar u \gamma_\mu \gamma_5 u_p
\\ \nonumber && ~~~~~~~~~~~~~~~~~~~ - {i c_1\over6}~ f^T_{J/\psi} f_p ~ q_1^\mu \epsilon_1^\nu \bar u \sigma_{\mu\nu} \gamma_5 u_p
\\ \nonumber &\equiv& C_{\psi p}~\epsilon_1^\mu \bar u \gamma_\mu \gamma_5 u_p + C^\prime_{\psi p}~q_1^\mu \epsilon_1^\nu \bar u \sigma_{\mu\nu} \gamma_5 u_p \, ,
\end{eqnarray}
where the two coupling constants $C_{\psi p}$ and $C^\prime_{\psi p}$ are defined for
\begin{eqnarray}
\mathcal{L}_{\psi p} &=& C_{\psi p}~\bar P_c \gamma_\mu \gamma_5 N~\psi^\mu \, ,
\label{lag:psipA}
\\ \mathcal{L}^\prime_{\psi p} &=& C^\prime_{\psi p}~\bar P_c \sigma_{\mu\nu} \gamma_5 N~\partial^\mu\psi^\nu \, .
\label{lag:psipB}
\end{eqnarray}

\item The decay of $|\bar D^{*0} \Sigma_c^{+}; 1/2^- \rangle$ into $\chi_{c0}(1P) p$ is contributed by $[\bar c_a c_a]~ \gamma_5 N$:
\begin{eqnarray}
&& \langle \bar D^{*0} \Sigma_c^{+}; 1/2^-(q) | \chi_{c0}(q_1)~p(q_2) \rangle
\\ \nonumber &\approx& {c_1\over6}~ m_{\chi_{c0}} f_{\chi_{c0}} f_p ~ \bar u \gamma_5 u_p
\\ \nonumber &\equiv& C_{\chi_{c0} p}~\bar u \gamma_5 u_p \, ,
\end{eqnarray}
where $C_{\chi_{c0} p}$ is defined for
\begin{eqnarray}
\mathcal{L}_{\chi_{c0} p} &=& C_{\chi_{c0} p}~\bar P_c \gamma_5 N~\chi_{c0} \, .
\end{eqnarray}

\item The decay of $|\bar D^{*0} \Sigma_c^{+}; 1/2^- \rangle$ into $\chi_{c1}(1P) p$ is contributed by $[\bar c_a \gamma_\mu \gamma_5 c_a]~\gamma^\mu N$:
\begin{eqnarray}
&& \langle \bar D^{*0} \Sigma_c^{+}; 1/2^-(q) | \chi_{c1}(q_1,\epsilon_1)~p(q_2) \rangle
\\ \nonumber &\approx& {c_1\over12}~ m_{\chi_{c1}} f_{\chi_{c1}} f_p ~ \epsilon_1^\mu \bar u \gamma_\mu u_p
\\ \nonumber &\equiv& C_{\chi_{c1} p}~\epsilon_1^\mu \bar u \gamma_\mu u_p \, ,
\end{eqnarray}
where $C_{\chi_{c1} p}$ is defined for
\begin{eqnarray}
\mathcal{L}_{\chi_{c1} p} &=& C_{\chi_{c1} p}~\bar P_c \gamma_\mu N~\chi_{c1}^\mu \, .
\end{eqnarray}
This decay channel may be kinematically allowed, depending on whether the $P_c(4457)$ is interpreted as $|\bar D^{*0} \Sigma_c^{+}; 1/2^- \rangle$ or not.

\end{enumerate}
Then we use the Fierz rearrangement given in Eq.~(\ref{eq:eta2etaxi}) to study the decay processes depicted in Fig.~\ref{fig:decayeta}(b,c):
\begin{enumerate}

\item[5.] The decay of $|\bar D^{*0} \Sigma_c^{+}; 1/2^- \rangle$ into $\bar D^{0} \Lambda_c^+$ is contributed by $[\bar c_a \gamma_5 u_a]~ \Lambda_c^+$:
\begin{eqnarray}
&& \langle \bar D^{*0} \Sigma_c^{+}; 1/2^-(q) ~|~\bar D^{0}(q_1)~\Lambda_c^+(q_2) \rangle
\\ \nonumber &\approx& -{ic_2\over3}~ \lambda_{D} f_{\Lambda_c} ~ \bar u u_{\Lambda_c}
\\ \nonumber &\equiv& C_{\bar D \Lambda_c}~\bar u u_{\Lambda_c} \, ,
\end{eqnarray}
where $c_2$ is an overall factor, and the coupling constant $C_{\bar D \Lambda_c}$ is defined for
\begin{eqnarray}
\mathcal{L}_{\bar D \Lambda_c} &=& C_{\bar D \Lambda_c}~\bar P_c \Lambda_c^+~\bar D^{0} \, .
\end{eqnarray}

\item[6.] The decay of $|\bar D^{*0} \Sigma_c^{+}; 1/2^- \rangle$ into $\bar D^{*0} \Lambda_c^+$ is contributed by $[\bar c_a \sigma_{\mu\nu} u_a] ~ \sigma^{\mu\nu} \gamma_5 \Lambda_c^+$:
\begin{eqnarray}
&& \langle \bar D^{*0} \Sigma_c^{+}; 1/2^-(q) ~|~\bar D^{*0}(q_1,\epsilon_1)~\Lambda_c^+(q_2) \rangle
\\ \nonumber &\approx& - {i c_2\over6}~ f^T_{D^*} f_{\Lambda_c} ~ q_1^\mu \epsilon_1^\nu \bar u \sigma_{\mu\nu} \gamma_5 u_{\Lambda_c}
\\ \nonumber &\equiv& C^\prime_{\bar D^{*} \Lambda_c}~q_1^\mu \epsilon_1^\nu \bar u \sigma_{\mu\nu} \gamma_5 u_{\Lambda_c} \, ,
\end{eqnarray}
where $C^\prime_{\bar D^{*} \Lambda_c}$ is defined for
\begin{eqnarray}
\mathcal{L}^\prime_{\bar D^{*} \Lambda_c} &=& C^\prime_{\bar D^{*} \Lambda_c}~\bar P_c \sigma_{\mu\nu} \gamma_5 \Lambda_c^+~\partial^\mu\bar D^{*0,\nu} \, .
\end{eqnarray}

\item[7.] Decays of $|\bar D^{*0} \Sigma_c^{+}; 1/2^- \rangle$ into the $\bar D^{0} \Sigma_c^+$ and $\bar D^{-} \Sigma_c^{++}$ final states are:
\begin{eqnarray}
&& \langle \bar D^{*0} \Sigma_c^{+}; 1/2^-(q) ~|~\bar D^{0}(q_1)~\Sigma_c^+(q_2) \rangle
\\ \nonumber &\approx& {ic_2\over8}~ f_{D} f_{\Sigma_c} ~ q_1^\mu \bar u \gamma_\mu u_{\Sigma_c}
\\ \nonumber &\equiv& C_{\bar D \Sigma_c}~ q_1^\mu \bar u \gamma_\mu u_{\Sigma_c} \, ,
\\ && \langle \bar D^{*0} \Sigma_c^{+}; 1/2^-(q) ~|~\bar D^{-}(q_1)~\Sigma_c^{++}(q_2) \rangle
\\ \nonumber &\approx& {i\sqrt2c_2\over8}~ f_{D} f_{\Sigma_c} ~ q_1^\mu \bar u \gamma_\mu u_{\Sigma_c}
\\ \nonumber &\equiv& \sqrt2C_{\bar D \Sigma_c}~q_1^\mu \bar u \gamma_\mu u_{\Sigma_c} \, ,
\end{eqnarray}
where $C_{\bar D \Sigma_c}$ is defined for
\begin{eqnarray}
\mathcal{L}_{\bar D \Sigma_c} &=& C_{\bar D \Sigma_c}~\bar P_c \gamma_\mu \Sigma_c^+~\partial^\mu\bar D^{0}
\\ \nonumber && ~~~~~~ + \sqrt2C_{\bar D \Sigma_c}~\bar P_c \gamma_\mu \Sigma_c^{++}~\partial^\mu\bar D^{-} \, .
\end{eqnarray}

\end{enumerate}

In the molecular picture the $P_c(4440)$ is sometimes interpreted as the $\bar D^* \Sigma_c$ hadronic molecular state of $J^P = {1/2}^-$. Accordingly, we assume the mass of $|\bar D^{*0} \Sigma_c^{+}; 1/2^- \rangle$ to be 4440.3~MeV, and summarize the above decay amplitudes to obtain the following (relative) decay widths:
\begin{eqnarray}
\nonumber \Gamma(|\bar D^{*0} \Sigma_c^{+}; 1/2^- \rangle \to \eta_c p ) &=& c_1^2 ~ 5.8 \times 10^{4}~{\rm GeV}^7  ,
\\[1mm] \nonumber \Gamma(|\bar D^{*0} \Sigma_c^{+}; 1/2^- \rangle \to J/\psi p ) &=& c_1^2 ~ 4.6 \times 10^{5}~{\rm GeV}^7  ,
\\[1mm] \nonumber \Gamma(|\bar D^{*0} \Sigma_c^{+}; 1/2^- \rangle \to \chi_{c0} p ) &=& c_1^2 ~ 2.0 \times 10^{3}~{\rm GeV}^7  ,
\\[1mm] \nonumber \Gamma(|\bar D^{*0} \Sigma_c^{+}; 1/2^- \rangle \to \bar D^{0} \Lambda_c^+ ) &=& c_2^2 ~ 5.5 \times 10^{5}~{\rm GeV}^7  ,
\\[1mm] \nonumber \Gamma(|\bar D^{*0} \Sigma_c^{+}; 1/2^- \rangle \to \bar D^{*0} \Lambda_c^+ ) &=& c_2^2 ~ 1.9 \times 10^{5}~{\rm GeV}^7  ,
\\[1mm] \nonumber \Gamma(|\bar D^{*0} \Sigma_c^{+}; 1/2^- \rangle \to \bar D^{0} \Sigma_c^+ ) &=& c_2^2 ~ 1.6 \times 10^{5}~{\rm GeV}^7  ,
\\[1mm] \nonumber \Gamma(|\bar D^{*0} \Sigma_c^{+}; 1/2^- \rangle \to \bar D^{-} \Sigma_c^{++} ) &=& c_2^2 ~ 3.2 \times 10^{5}~{\rm GeV}^7 .
\\ \label{result:eta2}
\end{eqnarray}
Besides, $|\bar D^{*0} \Sigma_c^{+}; 1/2^- \rangle$ can also couple to $\chi_{c1} p$, but this channel is kinematically forbidden under the assumption $M_{|\bar D^{*0} \Sigma_c^{+}; 1/2^- \rangle} = 4440.3$~MeV.

There are two different effective Lagrangians for the $|\bar D^{*0} \Sigma_c^{+}; 1/2^- \rangle$ decays into the $J/\psi p$ final state, as given in Eqs.~(\ref{lag:psipA}) and (\ref{lag:psipB}). It is interesting to see their individual contributions:
\begin{eqnarray}
\nonumber \Gamma(|\bar D^{*0} \Sigma_c^{+}; 1/2^- \rangle \to J/\psi p )\big|_{\mathcal{L}_{\psi p}} &=& c_1^2 ~ 1.5 \times 10^{5}~{\rm GeV}^7 ,
\\[1mm]
\nonumber \Gamma(|\bar D^{*0} \Sigma_c^{+}; 1/2^- \rangle \to J/\psi p )\big|_{\mathcal{L}^\prime_{\psi p}} &=& c_1^2 ~ 6.1 \times 10^{5}~{\rm GeV}^7 .
\\
\end{eqnarray}
Hence, the former is about four times smaller than the latter. Again, the phase angle between them can be important, whose relevant uncertainty will be investigated in Appendix~\ref{app:phase}.

\subsection{$\xi_2 \rightarrow \theta / \eta$}
\label{sec:decayxi2}

In this subsection we follow the procedures used in the previous subsection to study decay properties of $|\bar D^{*-} \Sigma_c^{++}; 1/2^- \rangle$, through the $\xi_2(x,y)$ current and the Fierz rearrangements given in Eqs.~(\ref{eq:xi2theta}) and (\ref{eq:xi2eta}). Again, we assume its mass to be 4440.3~MeV, and obtain the following (relative) decay widths:
\begin{eqnarray}
\nonumber \Gamma(|\bar D^{*-} \Sigma_c^{++}; 1/2^- \rangle \to \eta_c p ) &=& d_1^2 ~ 1.2 \times 10^{5}~{\rm GeV}^7  ,
\\[1mm] \nonumber \Gamma(|\bar D^{*-} \Sigma_c^{++}; 1/2^- \rangle \to J/\psi p ) &=& d_1^2 ~ 9.3 \times 10^{5}~{\rm GeV}^7  ,
\\[1mm] \nonumber \Gamma(|\bar D^{*-} \Sigma_c^{++}; 1/2^- \rangle \to \chi_{c0} p ) &=& d_1^2 ~ 4.1 \times 10^{3}~{\rm GeV}^7  ,
\\[1mm] \nonumber \Gamma(|\bar D^{*-} \Sigma_c^{++}; 1/2^- \rangle \to \bar D^{0} \Lambda_c^+ ) &=& d_2^2 ~ 1.1 \times 10^{6}~{\rm GeV}^7  ,
\\[1mm] \nonumber \Gamma(|\bar D^{*-} \Sigma_c^{++}; 1/2^- \rangle \to \bar D^{*0} \Lambda_c^+ ) &=& d_2^2 ~ 3.8 \times 10^{5}~{\rm GeV}^7  ,
\\[1mm] \nonumber \Gamma(|\bar D^{*-} \Sigma_c^{++}; 1/2^- \rangle \to \bar D^{0} \Sigma_c^+ ) &=& d_2^2 ~ 3.2 \times 10^{5}~{\rm GeV}^7  .
\\ \label{result:xi2}
\end{eqnarray}
Here $d_1$ and $d_2$ are two overall factors, which we simply assume to be $d_1 = c_1$ and $d_2 = c_2$ in the following analyses.

The above results suggest that $|\bar D^{*-} \Sigma_c^{++}; 1/2^- \rangle$ can not fall-apart decay into the $\bar D^{-} \Sigma_c^{++}$ final state, as depicted in Fig.~\ref{fig:decayxi}(b,c), while $|\bar D^{*0} \Sigma_c^{+}; 1/2^- \rangle$ can. The widths of the $|\bar D^{*-} \Sigma_c^{++}; 1/2^- \rangle$ decays into other final states, including $\eta_c p$, $J/\psi p$, $\chi_{c0} p$, $\bar D^{0} \Lambda_c^+$, $\bar D^{*0} \Lambda_c^+$, and $\bar D^{0} \Sigma_c^+$, are all two times larger than those given in Eqs.~(\ref{result:eta2}) for the $|\bar D^{*0} \Sigma_c^{+}; 1/2^- \rangle$ decays.

\subsection{$\eta_3^\alpha \rightarrow \theta / \eta / \xi$}
\label{sec:decayeta3}

In this subsection we follow the procedures used in Sec.~\ref{sec:decayeta1} and Sec.~\ref{sec:decayeta2} to study decay properties of $|\bar D^{*0} \Sigma_c^{+}; 3/2^- \rangle$ through the $\eta_3^\alpha(x,y)$ current. First we use the Fierz rearrangement given in Eq.~(\ref{eq:eta3theta}) to study the decay process depicted in Fig.~\ref{fig:decayeta}(a):
\begin{enumerate}

\item The decay of $|\bar D^{*0} \Sigma_c^{+}; 3/2^- \rangle$ into $\eta_c p$ is
\begin{eqnarray}
&& \langle \bar D^{*0} \Sigma_c^{+}; 3/2^-(q) ~|~\eta_c(q_1)~p(q_2) \rangle
\\ \nonumber &\approx& ie_1 ~ f_{\eta_c} f_p ~ q_1^\mu \bar u^\alpha \left({1\over16} g^{\alpha\mu} \gamma_5 + {i\over48} \sigma^{\alpha\mu} \gamma_5 \right) u_p \, ,
\end{eqnarray}
where $u^\alpha$ is the spinor of the $P_c$ state with $J^P = 3/2^-$, and $e_1$ is an overall factor.

\item The decay of $|\bar D^{*0} \Sigma_c^{+}; 3/2^- \rangle$ into $J/\psi p$ is
\begin{eqnarray}
&& \langle \bar D^{*0} \Sigma_c^{+}; 3/2^-(q) | J/\psi(q_1,\epsilon_1)~p(q_2) \rangle
\\ \nonumber &\approx& e_1 ~ m_{J/\psi} f_{J/\psi} f_p ~ \epsilon_1^\mu \bar u^\alpha \left(-{1\over16} g^{\alpha\mu} - {i\over48} \sigma^{\alpha\mu} \right) u_p \, .
\end{eqnarray}

\item The decay of $|\bar D^{*0} \Sigma_c^{+}; 3/2^- \rangle$ into $\chi_{c1}(1P) p$ is
\begin{eqnarray}
&& \langle \bar D^{*0} \Sigma_c^{+}; 3/2^-(q) | \chi_{c1}(q_1,\epsilon_1)~p(q_2) \rangle
\\ \nonumber &\approx& e_1 ~ m_{\chi_{c1}} f_{\chi_{c1}} f_p ~ \epsilon_1^\mu \bar u^\alpha \left({1\over16} g^{\alpha\mu} \gamma_5 + {i\over48} \sigma^{\alpha\mu} \gamma_5 \right) u_p \, .
\end{eqnarray}
This decay channel may be kinematically allowed, depending on whether the $P_c(4457)$ is interpreted as $|\bar D^{*0} \Sigma_c^{+}; 3/2^- \rangle$ or not.

\end{enumerate}
Then we use the Fierz rearrangement given in Eq.~(\ref{eq:eta3etaxi}) to study the decay processes depicted in Fig.~\ref{fig:decayeta}(b,c):
\begin{enumerate}

\item[4.] The decay of $|\bar D^{*0} \Sigma_c^{+}; 3/2^- \rangle$ into $\bar D^{*0} \Lambda_c^+$ is
\begin{eqnarray}
&& \langle \bar D^{*0} \Sigma_c^{+}; 3/2^-(q) ~|~\bar D^{*0}(q_1,\epsilon_1)~\Lambda_c^+(q_2) \rangle
\\ \nonumber &\approx& 2i e_2 ~ f^T_{D^*} f_{\Lambda_c} ~ q_1^\mu \epsilon_1^\nu ~ \times ~
\\ \nonumber && ~ \bar u^\alpha \left(- {i\over48} g^{\alpha\mu} \gamma^\nu + {i\over48} g^{\alpha\nu} \gamma^\mu - {1\over48} \epsilon^{\alpha\beta\mu\nu} \gamma_\beta \gamma_5 \right) u_{\Lambda_c},
\end{eqnarray}
where $e_2$ is an overall factor.

\item[5.] Decays of $|\bar D^{*0} \Sigma_c^{+}; 3/2^- \rangle$ into the $\bar D^{0} \Sigma_c^+$ and $\bar D^{-} \Sigma_c^{++}$ final states are:
\begin{eqnarray}
&& \langle \bar D^{*0} \Sigma_c^{+}; 3/2^-(q) ~|~\bar D^{0}(q_1)~\Sigma_c^+(q_2) \rangle
\\ \nonumber &\approx& i e_2 ~ f_{D} f_{\Sigma_c} ~ q_1^\mu
\\ \nonumber && \times ~ \bar u^\alpha \left({1\over24} g^{\alpha\mu} \gamma^\nu - {1\over24} g^{\alpha\nu} \gamma^\mu - {i\over24} \epsilon^{\alpha\beta\mu\nu} \gamma_\beta \gamma_5 \right)
\\ \nonumber && ~~~~~~~~~~~~~~~~~~~~~~~~~~~~~~~~~~~~~~~ \times \left( - {1\over4} \gamma_\nu \gamma_5 \right) u_{\Sigma_c} \, ,
\\ && \langle \bar D^{*0} \Sigma_c^{+}; 3/2^-(q) ~|~\bar D^{-}(q_1)~\Sigma_c^{++}(q_2) \rangle
\\ \nonumber &\approx& \sqrt2 i e_2 ~ f_{D} f_{\Sigma_c} ~ q_1^\mu
\\ \nonumber && \times ~ \bar u^\alpha \left({1\over24} g^{\alpha\mu} \gamma^\nu - {1\over24} g^{\alpha\nu} \gamma^\mu - {i\over24} \epsilon^{\alpha\beta\mu\nu} \gamma_\beta \gamma_5 \right)
\\ \nonumber && ~~~~~~~~~~~~~~~~~~~~~~~~~~~~~~~~~~~~~~ \times \left( - {1\over4} \gamma_\nu \gamma_5 \right) ~ u_{\Sigma_c} \, .
\end{eqnarray}

\end{enumerate}

In the molecular picture the $P_c(4457)$ is sometimes interpreted as the $\bar D^* \Sigma_c$ hadronic molecular state of $J^P = {3/2}^-$. Accordingly, we assume the mass of $|\bar D^{*0} \Sigma_c^{+}; 3/2^- \rangle$ to be 4457.3~MeV, and summarize the above decay amplitudes to obtain the following (relative) decay widths:
\begin{eqnarray}
\nonumber \Gamma(|\bar D^{*0} \Sigma_c^{+}; 3/2^- \rangle \to \eta_c p ) &=& e_1^2 ~ 240~{\rm GeV}^7  ,
\\[1mm] \nonumber \Gamma(|\bar D^{*0} \Sigma_c^{+}; 3/2^- \rangle \to J/\psi p ) &=& e_1^2 ~ 4.7 \times 10^{4}~{\rm GeV}^7  ,
\\[1mm] \nonumber \Gamma(|\bar D^{*0} \Sigma_c^{+}; 3/2^- \rangle \to \chi_{c1} p ) &=& e_1^2 ~ 15~{\rm GeV}^7  ,
\\[1mm] \nonumber \Gamma(|\bar D^{*0} \Sigma_c^{+}; 3/2^- \rangle \to \bar D^{*0} \Lambda_c^+ ) &=& e_2^2 ~ 1.6 \times 10^{4}~{\rm GeV}^7  ,
\\[1mm] \nonumber \Gamma(|\bar D^{*0} \Sigma_c^{+}; 3/2^- \rangle \to \bar D^{0} \Sigma_c^+ ) &=& e_2^2 ~ 5.7 ~{\rm GeV}^7  ,
\\[1mm] \nonumber \Gamma(|\bar D^{*0} \Sigma_c^{+}; 3/2^- \rangle \to \bar D^{-} \Sigma_c^{++} ) &=& e_2^2 ~ 11~{\rm GeV}^7 .
\\ \label{result:eta3}
\end{eqnarray}
Hence, $|\bar D^{*0} \Sigma_c^{+}; 3/2^- \rangle$ does not couple to the $\chi_{c0} p$ channel, different from $|\bar D^{*0} \Sigma_c^{+}; 1/2^- \rangle$.

\subsection{$\xi_3^\alpha \rightarrow \theta / \eta$}
\label{sec:decayxi3}

In this subsection we follow the procedures used in the previous subsection to study decay properties of $|\bar D^{*-} \Sigma_c^{++}; 3/2^- \rangle$, through the $\xi_3^\alpha(x,y)$ current and the Fierz rearrangements given in Eqs.~(\ref{eq:xi3theta}) and (\ref{eq:xi3eta}). Again, we assume its mass to be 4457.3~MeV, and obtain the following (relative) decay widths:
\begin{eqnarray}
\nonumber \Gamma(|\bar D^{*-} \Sigma_c^{++}; 3/2^- \rangle \to \eta_c p ) &=& f_1^2 ~ 490~{\rm GeV}^7  ,
\\[1mm] \nonumber \Gamma(|\bar D^{*-} \Sigma_c^{++}; 3/2^- \rangle \to J/\psi p ) &=& f_1^2 ~ 9.3 \times 10^{4}~{\rm GeV}^7  ,
\\[1mm] \nonumber \Gamma(|\bar D^{*-} \Sigma_c^{++}; 3/2^- \rangle \to \chi_{c1} p ) &=& f_1^2 ~ 30~{\rm GeV}^7  ,
\\[1mm] \nonumber \Gamma(|\bar D^{*-} \Sigma_c^{++}; 3/2^- \rangle \to \bar D^{*0} \Lambda_c^+ ) &=& f_2^2 ~ 3.3 \times 10^{4}~{\rm GeV}^7  ,
\\[1mm] \nonumber \Gamma(|\bar D^{*-} \Sigma_c^{++}; 3/2^- \rangle \to \bar D^{0} \Sigma_c^+ ) &=& f_2^2 ~ 11 ~{\rm GeV}^7  .
\\ \label{result:xi3}
\end{eqnarray}
Here $f_1$ and $f_2$ are two overall factors, which we simply assume to be $f_1 = e_1$ and $f_2 = e_2$ in the following analyses.

The above results suggest that $|\bar D^{*-} \Sigma_c^{++}; 3/2^- \rangle$ can not fall-apart decay into the $\bar D^{-} \Sigma_c^{++}$ final state, as depicted in Fig.~\ref{fig:decayxi}(b,c), while $|\bar D^{*0} \Sigma_c^{+}; 3/2^- \rangle$ can. The widths of the $|\bar D^{*-} \Sigma_c^{++}; 3/2^- \rangle$ decays into other final states, including $\eta_c p$, $J/\psi p$, $\chi_{c1} p$, $\bar D^{*0} \Lambda_c^+$, and $\bar D^{0} \Sigma_c^+$, are all two times larger than those given in Eqs.~(\ref{result:eta3}) for the $|\bar D^{*0} \Sigma_c^{+}; 3/2^- \rangle$ decays.

\section{Isospin of $\bar D^{(*)} \Sigma_c$ molecular states}
\label{sec:isospin}

In this section we collect the results calculated in the previous section to further study decay properties of $\bar D^{(*)} \Sigma_c$ molecular states with definite isospins.

The $\bar D^{(*)} \Sigma_c$ molecular states with $I = 1/2$ can be obtained by using Eqs.~(\ref{def:molecule1}), (\ref{def:molecule2}), and (\ref{def:molecule3}) with $\theta_i = -55^{\rm o}$:
\begin{eqnarray}
&& | \bar D^{(*)} \Sigma_c; {1\over2}^-/{3\over2}^- \rangle
\label{isospin:1half}
\\ \nonumber && ~~ = \sqrt{1\over3}~| \bar D^{(*)0} \Sigma_c^+ \rangle_{J={1\over2}/{3\over2}} - \sqrt{2\over3}~| \bar D^{(*)-} \Sigma_c^{++} \rangle_{J={1\over2}/{3\over2}} \, .
\end{eqnarray}
Combining the results of Sec.~\ref{sec:decayeta1} and Sec.~\ref{sec:decayxi1}, we obtain:
\begin{eqnarray}
\nonumber \Gamma(|\bar D \Sigma_c; 1/2^- \rangle \to \eta_c p ) &=& a_1^2 ~ 3.2 \times 10^{5}~{\rm GeV}^7 \, ,
\\[1mm] \nonumber \Gamma(|\bar D \Sigma_c; 1/2^- \rangle \to J/\psi p ) &=& a_1^2 ~ 8.5 \times 10^{4}~{\rm GeV}^7 \, ,
\\[1mm] \nonumber \Gamma(|\bar D \Sigma_c; 1/2^- \rangle \to \bar D^{*0} \Lambda_c^+ ) &=& a_2^2 ~ 5.9 \times 10^{4}~{\rm GeV}^7 \, .
\\ \label{result:etaxi1}
\end{eqnarray}
Combining the results of Sec.~\ref{sec:decayeta2} and Sec.~\ref{sec:decayxi2}, we obtain:
\begin{eqnarray}
\nonumber \Gamma(|\bar D^{*} \Sigma_c; 1/2^- \rangle \to \eta_c p ) &=& c_1^2 ~ 1.7 \times 10^{5}~{\rm GeV}^7  ,
\\[1mm] \nonumber \Gamma(|\bar D^{*} \Sigma_c; 1/2^- \rangle \to J/\psi p ) &=& c_1^2 ~ 1.4 \times 10^{6}~{\rm GeV}^7  ,
\\[1mm] \nonumber \Gamma(|\bar D^{*} \Sigma_c; 1/2^- \rangle \to \chi_{c0} p ) &=& c_1^2 ~ 6.1 \times 10^{3}~{\rm GeV}^7  ,
\\[1mm] \nonumber \Gamma(|\bar D^{*} \Sigma_c; 1/2^- \rangle \to \bar D^{0} \Lambda_c^+ ) &=& c_2^2 ~ 1.7 \times 10^{6}~{\rm GeV}^7  ,
\\[1mm] \nonumber \Gamma(|\bar D^{*} \Sigma_c; 1/2^- \rangle \to \bar D^{*0} \Lambda_c^+ ) &=& c_2^2 ~ 5.6 \times 10^{5}~{\rm GeV}^7  ,
\\[1mm] \nonumber \Gamma(|\bar D^{*} \Sigma_c; 1/2^- \rangle \to \bar D^{0} \Sigma_c^+ ) &=& c_2^2 ~ 5.4 \times 10^{4}~{\rm GeV}^7  ,
\\[1mm] \nonumber \Gamma(|\bar D^{*} \Sigma_c; 1/2^- \rangle \to \bar D^{-} \Sigma_c^{++} ) &=& c_2^2 ~ 1.1 \times 10^{5}~{\rm GeV}^7 .
\\ \label{result:etaxi2}
\end{eqnarray}
Combining the results of Sec.~\ref{sec:decayeta3} and Sec.~\ref{sec:decayxi3}, we obtain:
\begin{eqnarray}
\nonumber \Gamma(|\bar D^{*} \Sigma_c; 3/2^- \rangle \to \eta_c p ) &=& e_1^2 ~ 730~{\rm GeV}^7  ,
\\[1mm] \nonumber \Gamma(|\bar D^{*} \Sigma_c; 3/2^- \rangle \to J/\psi p ) &=& e_1^2 ~ 1.4 \times 10^{5}~{\rm GeV}^7  ,
\\[1mm] \nonumber \Gamma(|\bar D^{*} \Sigma_c; 3/2^- \rangle \to \chi_{c1} p ) &=& e_1^2 ~ 46~{\rm GeV}^7  ,
\\[1mm] \nonumber \Gamma(|\bar D^{*} \Sigma_c; 3/2^- \rangle \to \bar D^{*0} \Lambda_c^+ ) &=& e_2^2 ~ 4.9 \times 10^{4}~{\rm GeV}^7  ,
\\[1mm] \nonumber \Gamma(|\bar D^{*} \Sigma_c; 3/2^- \rangle \to \bar D^{0} \Sigma_c^+ ) &=& e_2^2 ~ 1.9 ~{\rm GeV}^7  ,
\\[1mm] \nonumber \Gamma(|\bar D^{*} \Sigma_c; 3/2^- \rangle \to \bar D^{-} \Sigma_c^{++} ) &=& e_2^2 ~ 3.8 ~{\rm GeV}^7  .
\\ \label{result:etaxi3}
\end{eqnarray}
Comparing the above values with those given in Eqs.~(\ref{result:eta1}), (\ref{result:eta2}), and (\ref{result:eta3}), we find that the decay widths of the three $\bar D^{(*)} \Sigma_c$ molecular states with $I = 1/2$ into the $\eta_c p$, $J/\psi p$, $\chi_{c0} p$, $\chi_{c1} p$, $\bar D^{0} \Lambda_c^+$, and $\bar D^{*0} \Lambda_c^+$ final states also with $I = 1/2$ are increased by three times, and their decay widths into the $\bar D^{0} \Sigma_c^+$ and $\bar D^{-} \Sigma_c^{++}$ final states are decreased by three times. We shall further discuss these results in Sec.~\ref{sec:summary}.

For completeness, we also list here the results for the three $\bar D^{(*)} \Sigma_c$ molecular states with $I = 3/2$ (as if they existed), which can be obtained by using Eqs.~(\ref{def:molecule1}), (\ref{def:molecule2}), and (\ref{def:molecule3}) with $\theta_i = 35^{\rm o}$:
\begin{eqnarray}
&& | \bar D^{(*)} \Sigma_c; {1\over2}^{-\prime}/{3\over2}^{-\prime} \rangle
\label{isospin:3half}
\\ \nonumber && ~~ = \sqrt{2\over3}~| \bar D^{(*)0} \Sigma_c^+ \rangle_{J={1\over2}/{3\over2}} + \sqrt{1\over3}~| \bar D^{(*)-} \Sigma_c^{++} \rangle_{J={1\over2}/{3\over2}} \, .
\end{eqnarray}
Naively assuming their masses to be $4311.9$~MeV, $4440.3$~MeV, and $4457.3$~MeV, respectively, we obtain the following non-zero (relative) decay widths:
\begin{eqnarray}
\nonumber \Gamma(|\bar D^{*} \Sigma_c; 1/2^{-\prime} \rangle \to \bar D^{0} \Sigma_c^+ ) &=& c_2^2 ~ 4.3 \times 10^{5}~{\rm GeV}^7  ,
\\[1mm] \nonumber \Gamma(|\bar D^{*} \Sigma_c; 1/2^{-\prime} \rangle \to \bar D^{-} \Sigma_c^{++} ) &=& c_2^2 ~ 2.2 \times 10^{5}~{\rm GeV}^7  ,
\\[1mm] \nonumber \Gamma(|\bar D^{*} \Sigma_c; 3/2^{-\prime} \rangle \to \bar D^{0} \Sigma_c^+ ) &=& e_2^2 ~ 15~{\rm GeV}^7  ,
\\[1mm] \nonumber \Gamma(|\bar D^{*} \Sigma_c; 3/2^{-\prime} \rangle \to \bar D^{-} \Sigma_c^{++} ) &=& e_2^2 ~ 7.6 ~{\rm GeV}^7  .
\\ \label{result:isospin3half}
\end{eqnarray}
Comparing them with Eqs.~(\ref{result:eta1}), (\ref{result:eta2}), and (\ref{result:eta3}), we find that the three $\bar D^{(*)} \Sigma_c$ molecular states with $I = 3/2$ can not fall-apart decay into the $\eta_c p$, $J/\psi p$, $\chi_{c0} p$, $\chi_{c1} p$, $\bar D^{0} \Lambda_c^+$, and $\bar D^{*0} \Lambda_c^+$ final states with $I = 1/2$, their widths into the $\bar D^{0} \Sigma_c^+$ final state are increased by a factor of $8/3$, and their widths into the $\bar D^{-} \Sigma_c^{++}$ final state are reduced to two third. We summarize these results in Appendix~\ref{app:phase}, which we shall not discuss any more.

\section{Summary and conclusions}
\label{sec:summary}

In this paper we systematically study hidden-charm pentaquark currents with the quark content $\bar c c u u d$. We investigate three different configurations, $\eta = [\bar c u][u d c]$, $\xi = [\bar c d][u u c]$, and $\theta = [\bar c c][u u d]$. Some of their relations are derived using the Fierz rearrangement of the Dirac and color indices, and the obtained results are used to study strong decay properties of $\bar D^{(*)} \Sigma_c$ molecular states with $I = 1/2$ and $J^P = 1/2^-$ and $3/2^-$.

Before drawing conclusions, we would like to generally discuss about the uncertainty. In the present study we work under the naive factorization scheme, so our uncertainty is larger than the well-developed QCD factorization scheme~\cite{Beneke:1999br,Beneke:2000ry,Beneke:2001ev}, that is at the 5\% level when being applied to conventional (heavy) hadrons~\cite{Li:2020rcg}. On the other hand, the pentaquark decay constants, such as $f_{P_c}$, are removed when calculating relative branching ratios. This significantly reduces our uncertainty. Accordingly, we roughly estimate our uncertainty to be at the $X^{+100\%}_{-~50\%}$ level.

In the molecular picture the $P_c(4312)$ is usually interpreted as the $\bar D \Sigma_c$ hadronic molecular state of $J^P = {1/2}^-$, and the $P_c(4440)$ and $P_c(4457)$ are sometimes interpreted as the $\bar D^* \Sigma_c$ hadronic molecular states of $J^P = {1/2}^-$ and ${3/2}^-$ respectively (sometimes interpreted as states of $J^P = {3/2}^-$ and ${1/2}^-$ respectively)~\cite{Wu:2012md,Chen:2019asm,Liu:2019tjn}. Using their masses measured in the LHCb experiment~\cite{Aaij:2019vzc} as inputs, we calculate some of their relative decay widths. The obtained results have been summarized in Eqs.~(\ref{result:etaxi1}), (\ref{result:etaxi2}), and (\ref{result:etaxi3}), from which we further obtain:

\begin{widetext}
\begin{itemize}

\item We obtain the following relative branching ratios for the $|\bar D \Sigma_c; 1/2^- \rangle$ decays:
\begin{eqnarray}
&& {\mathcal{B}\left(|\bar D \Sigma_c; 1/2^- \rangle \rightarrow
~~~J/\psi p~~~
: ~~~~~~~~\eta_c p~~~~~~~~
: ~~~\bar D^{*0} \Lambda_c^+~
\right) \over \mathcal{B}\left(|\bar D \Sigma_c; 1/2^- \rangle \rightarrow J/\psi p\right)}
\\[2mm] \nonumber &\approx&
~~~~~~~~~~~~~~~~~~~~~~~~~~~~~~~ 1 ~~~~~ : \,~~~~~~~~3.8~~~~~~~~\, : ~~~~0.69t~ \, .
\end{eqnarray}

\item We obtain the following relative branching ratios for the $|\bar D^* \Sigma_c; 1/2^- \rangle$ decays:
\begin{eqnarray}
&& {\mathcal{B}\left(|\bar D^* \Sigma_c; 1/2^- \rangle \rightarrow
~~J/\psi p~~
: ~~\eta_c p~~
: \,~\chi_{c0} p~\,
: ~\bar D^{0} \Lambda_c^+~
: ~\bar D^{*0} \Lambda_c^+~
: ~\bar D^{0} \Sigma_c^+~
: ~\bar D^{-} \Sigma_c^{++}~
\right) \over \mathcal{B}\left(|\bar D^* \Sigma_c; 1/2^- \rangle \rightarrow J/\psi p\right)}
\\[2mm] \nonumber &\approx&
~~~~~~~~~~~~~~~~~~~~~~~~~~~~~~~ 1 ~~~~\, : \,~0.13~\, : ~0.004~ : \,~~1.2t~~\, : \,~~0.41t~~\, : ~~0.04t~~ : ~~~0.08t~ \, .
\end{eqnarray}

\item We obtain the following relative branching ratios for the $|\bar D^* \Sigma_c; 3/2^- \rangle$ decays:
\begin{eqnarray}
&& {\mathcal{B}\left(|\bar D^* \Sigma_c; 3/2^- \rangle \rightarrow
~~J/\psi p~~
: ~~\eta_c p~~
: \,~\chi_{c1} p~\,
: ~\bar D^{*0} \Lambda_c^+~
: ~\bar D^{0} \Sigma_c^+~
: ~\bar D^{-} \Sigma_c^{++}~
\right) \over \mathcal{B}\left(|\bar D^* \Sigma_c; 3/2^- \rangle \rightarrow J/\psi p\right)}
\\[2mm] \nonumber &\approx&
~~~~~~~~~~~~~~~~~~~~~~~~~~~~~~~ 1 ~~~~\, : ~0.005~ : ~10^{-4}~ : \,~~0.35t~~\, : \,~10^{-5}t~\, : ~~~10^{-5}t~ \, .
\end{eqnarray}

\end{itemize}
\end{widetext}

In these expressions, $t \equiv {a_2^2 \over a_1^2} \approx {c_2^2 \over c_1^2} \approx {e_2^2 \over e_1^2}$ is the parameter measuring which processes happen more easily, the processes depicted in Figs.~\ref{fig:decayeta}\&\ref{fig:decayxi}(a) or the processes depicted in Figs.~\ref{fig:decayeta}\&\ref{fig:decayxi}(b,c). Generally speaking, the exchange of one light quark with another light quark seems to be easier than the exchange of one light quark with another heavy quark~\cite{Landau}, so it can be the case that $t \geq 1$.
There are two phase angles, which have not been taken into account in the above expressions yet. We investigate their relevant uncertainties in Appendix~\ref{app:phase}, where we also give the relative branching ratios for the $\bar D^{(*)} \Sigma_c$ hadronic molecular states of $I = 3/2$, and separately for the $\bar D^{(*)0} \Sigma_c^+$ and $\bar D^{(*)-} \Sigma_c^{++}$ hadronic molecular states.

To extract these results:
\begin{itemize}

\item We have only considered the leading-order fall-apart decays described by color-singlet-color-singlet meson-baryon currents, but neglected the $\mathcal{O}(\alpha_s)$ corrections described by color-octet-color-octet meson-baryon currents, so there can be other possible decay channels.

\item We have omitted all the charmed baryon fields of $J=3/2$, so we can not study decays of $P_c$ states into the $\bar D \Sigma_c^*$ final state. However, we have kept all the charmed baryon fields that can couple to the $J^P=1/2^+$ ground-state charmed baryons $\Lambda_c$ and $\Sigma_c$, {\it i.e.}, fields given in Eqs.~(\ref{eq:heavybaryon}), so decays of $P_c$ states into the $\bar D^{(*)} \Lambda_c$ and $\bar D \Sigma_c$ final states have been well investigated in the present study.

\item We have omitted all the light baryon fields of $J=3/2$, so we can not study decays of $P_c$ states into charmonia and $\Delta/N^*$. However, we have kept all the light baryon fields of $J^P=1/2^+$, {\it i.e.}, terms depending on $N_1$ and $N_2$, so decays of $P_c$ states into charmonia and protons have been well investigated in the present study.

\end{itemize}

Our conclusions are:
\begin{itemize}

\item Firstly, we compare the $\eta_c p$ and $J/\psi p$ channels:
\begin{eqnarray}
\nonumber    {\mathcal{B}\left(|\bar D \Sigma_c; 1/2^- \rangle \rightarrow \eta_c p \right) \over \mathcal{B}\left(|\bar D \Sigma_c; 1/2^- \rangle \rightarrow J/\psi p\right)} &\approx& 3.8 \, ,
\\           {\mathcal{B}\left(|\bar D^* \Sigma_c; 1/2^- \rangle \rightarrow \eta_c p \right) \over \mathcal{B}\left(|\bar D^* \Sigma_c; 1/2^- \rangle \rightarrow J/\psi p\right)} &\approx& 0.13 \, ,
\\ \nonumber {\mathcal{B}\left(|\bar D^* \Sigma_c; 3/2^- \rangle \rightarrow \eta_c p \right) \over \mathcal{B}\left(|\bar D^* \Sigma_c; 3/2^- \rangle \rightarrow J/\psi p\right)} &\approx& 0.005 \, .
\end{eqnarray}
These ratios are quite similar to those obtained using the heavy quark spin symmetry~\cite{Voloshin:2019aut}. This is quite reasonable because no spin symmetry breaking is introduced during the calculation before using the decay constants for the mesons, so that the heavy quark spin symmetry is automatically built in our formalism. Since the width of the $|\bar D \Sigma_c; 1/2^- \rangle$ decay into the $\eta_c p$ final state is comparable to its decay width into $J/\psi p$, we propose to confirm the existence of the $P_c(4312)$ in the $\eta_c p$ channel.

\item Secondly, we compare the $\bar D^{(*)} \Lambda_c$ and $J/\psi p$ channels:
\begin{eqnarray}
{\mathcal{B}\left(|\bar D^* \Sigma_c; 1/2^- \rangle \rightarrow \bar D^{0} \Lambda_c^+ \right) \over \mathcal{B}\left(|\bar D^* \Sigma_c; 1/2^- \rangle \rightarrow J/\psi p\right)} &\approx& 1.2t \, ,
\end{eqnarray}
and
\begin{eqnarray}
\nonumber {\mathcal{B}\left(|\bar D \Sigma_c; 1/2^- \rangle \rightarrow \bar D^{*0} \Lambda_c^+ \right) \over \mathcal{B}\left(|\bar D \Sigma_c; 1/2^- \rangle \rightarrow J/\psi p\right)} &\approx& 0.69t \, ,
\\ {\mathcal{B}\left(|\bar D^* \Sigma_c; 1/2^- \rangle \rightarrow \bar D^{*0} \Lambda_c^+ \right) \over \mathcal{B}\left(|\bar D^* \Sigma_c; 1/2^- \rangle \rightarrow J/\psi p\right)} &\approx& 0.41t \, ,
\\ \nonumber {\mathcal{B}\left(|\bar D^* \Sigma_c; 3/2^- \rangle \rightarrow \bar D^{*0} \Lambda_c^+ \right) \over \mathcal{B}\left(|\bar D^* \Sigma_c; 3/2^- \rangle \rightarrow J/\psi p\right)} &\approx& 0.35t \, .
\end{eqnarray}
Accordingly, we propose to observe the $P_c(4312)$, $P_c(4440)$, and $P_c(4457)$ in the $\bar D^{*0} \Lambda_c^+$ channel. Moreover, the $\bar D^{0} \Lambda_c^+$ channel can be an ideal channel to extract the spin-parity quantum numbers of the $P_c(4440)$ and $P_c(4457)$.

\item Thirdly, we compare the $\bar D \Sigma_c$ and $J/\psi p$ channels:
\begin{eqnarray}
{\mathcal{B}\left(|\bar D^* \Sigma_c; 1/2^- \rangle \rightarrow \bar D^{0} \Sigma_c^+ \right) \over \mathcal{B}\left(|\bar D^* \Sigma_c; 1/2^- \rangle \rightarrow J/\psi p\right)} &\approx& 0.04t \, ,
\\
\nonumber {\mathcal{B}\left(|\bar D^* \Sigma_c; 1/2^- \rangle \rightarrow \bar D^{-} \Sigma_c^{++} \right) \over \mathcal{B}\left(|\bar D^* \Sigma_c; 1/2^- \rangle \rightarrow J/\psi p\right)} &\approx& 0.08t \, ,
\end{eqnarray}
and
\begin{eqnarray}
{\mathcal{B}\left(|\bar D^* \Sigma_c; 3/2^- \rangle \rightarrow \bar D^{0} \Sigma_c^+ \right) \over \mathcal{B}\left(|\bar D^* \Sigma_c; 3/2^- \rangle \rightarrow J/\psi p\right)} &\approx& 10^{-5}t \, ,
\\
\nonumber {\mathcal{B}\left(|\bar D^* \Sigma_c; 3/2^- \rangle \rightarrow \bar D^{-} \Sigma_c^{++} \right) \over \mathcal{B}\left(|\bar D^* \Sigma_c; 3/2^- \rangle \rightarrow J/\psi p\right)} &\approx& 10^{-5}t \, .
\end{eqnarray}
Accordingly, we propose to observe the $P_c(4440)$ and $P_c(4457)$ in the $\bar D^{-} \Sigma_c^{++}$ channel, which is another possible channel to extract their spin-parity quantum numbers.

\end{itemize}

\section*{Acknowledgments}

We thank Utku Can, Philipp Gubler, and Makoto Oka for helpful discussions.
This project is supported by the National Natural Science Foundation of China under Grants No.~11722540 and No.~12075019.

\appendix

\section{Parameters and decay formulae}
\label{app:parameters}

We list masses of $P_c$ states used in the present study, taken from the LHCb experiment~\cite{Aaij:2019vzc}:
\begin{eqnarray}
   \nonumber        P_c(4312)^+  ~:~ m&=&4311.9 \mbox{ MeV} \, ,
\\         P_c(4440)^+  ~:~ m&=&4440.3 \mbox{ MeV} \, ,
\\ \nonumber        P_c(4457)^+  ~:~ m&=&4457.3 \mbox{ MeV} \, .
\end{eqnarray}
We list masses of charmonium mesons and charmed mesons used in the present study, taken from PDG~\cite{pdg} and partly averaged over isospin:
\begin{eqnarray}
   \nonumber        \eta_c(1S)     ~:~ m&=&2983.9 \mbox{ MeV} \, ,
\\ \nonumber        J/\psi(1S)     ~:~ m&=&3096.900 \mbox{ MeV} \, ,
\\         \chi_{c0}(1P)  ~:~ m&=&3414.71 \mbox{ MeV} \, ,
\\ \nonumber        \chi_{c1}(1P)  ~:~ m&=&3510.67 \mbox{ MeV} \, ,
\\ \nonumber        D/\bar D       ~:~ m&=&1867.24 \mbox{ MeV} \, ,
\\ \nonumber        D^*/\bar D^*   ~:~ m&=&2008.55 \mbox{ MeV} \, .
\end{eqnarray}
We list masses of the proton and charmed baryons used in the present study, taken from PDG~\cite{pdg} and partly averaged over isospin:
\begin{eqnarray}
   \nonumber        {\rm proton}         ~:~ m&=&938.272 \mbox{ MeV} \, ,
\\         \Lambda_c^+    ~:~ m&=&2286.46 \mbox{ MeV} \, ,
\\ \nonumber        \Sigma_c       ~:~ m&=&2453.44 \mbox{ MeV} \, .
\end{eqnarray}

In this paper we only investigate two-body decays, and their widths can be easily calculated. In the calculations we use the following formula for baryon fields of spin 1/2 and 3/2:
\begin{eqnarray}
\sum_{spin} u(p) \bar u(p) &=& \left( p\!\!\!\slash + m \right) \, ,
\\
\sum_{spin} u_{\mu}(p) \bar u_{\mu^\prime}(p) &=& \Big( g_{\mu\mu^\prime} - {1\over3} \gamma_\mu \gamma_{\mu^\prime}
\\ \nonumber &-& {p_\mu\gamma_{\mu^\prime} - p_{\mu^\prime}\gamma_{\mu} \over 3m} - {2p_{\mu}p_{\mu^\prime} \over 3m^2} \Big) \left( p\!\!\!\slash + m \right) \, .
\end{eqnarray}

\section{Heavy and light baryon fields}
\label{app:baryon}

First we construct charmed baryon interpolating fields. We refer to Ref.~\cite{Dmitrasinovic} for detailed discussions. There are altogether nine independent charmed baryon fields:
\begin{eqnarray}
B^G_{\mathbf{\bar 3},1} &=& \epsilon_{abc} \epsilon^{ABG} (q_A^{aT} \mathbb{C} q_B^b) \gamma_5 c^c \, ,
\label{def:heavy1}
\\
B^G_{\mathbf{\bar 3},2} &=& \epsilon_{abc} \epsilon^{ABG} (q_A^{aT} \mathbb{C} \gamma_5 q_B^b) c^c \, ,
\\
B^G_{\mathbf{\bar 3},3} &=& \epsilon_{abc} \epsilon^{ABG} (q_A^{aT} \mathbb{C} \gamma_\mu \gamma_5 q_B^b) \gamma^\mu c^c \, ,
\\
B^G_{\mathbf{\bar 3},\mu} &=& P^{3/2}_{\mu\nu} \epsilon_{abc} \epsilon^{ABG} (q_A^{aT} \mathbb{C} \gamma^\nu \gamma_5 q_B^b) \gamma_5 c^c \, ,
\\
B^U_{\mathbf{6},4} &=& \epsilon_{abc} S^U_{AB} (q_A^{aT} \mathbb{C} \gamma_\mu q_B^b) \gamma^\mu \gamma_5 c^c  \, ,
\\
B^U_{\mathbf{6},5} &=& \epsilon_{abc} S^U_{AB} (q_A^{aT} \mathbb{C} \sigma_{\mu\nu} q_B^b) \sigma^{\mu\nu} \gamma_5 c^c  \, ,
\\
B^U_{\mathbf{6},\mu} &=& P^{3/2}_{\mu\nu} \epsilon_{abc} S^U_{AB} (q_A^{aT} \mathbb{C} \gamma^\nu q_B^b) c^c \, ,
\\
B^{\prime U}_{\mathbf{6},\mu} &=& P^{3/2}_{\mu\nu} ( B^{U,\nu}_{\mathbf{6},7} + B^{U,\nu}_{\mathbf{6},8\nu}) \, ,
\\
B^U_{\mathbf{6},\mu\nu} &=&  P^{3/2}_{\mu\nu\alpha\beta} ( B^{U,\alpha\beta}_{\mathbf{6},7} + B^{U,\alpha\beta}_{\mathbf{6},8} ) \, ,
\label{def:heavy9}
\end{eqnarray}
where
\begin{eqnarray}
B^U_{\mathbf{6},7\mu} &=& \epsilon_{abc} S^U_{AB} (q_A^{aT} \mathbb{C} \sigma_{\mu\nu} q_B^b) \gamma^\nu c^c \, ,
\\
B^U_{\mathbf{6},8\mu} &=& \epsilon_{abc} S^U_{AB} (q_A^{aT} \mathbb{C} \sigma_{\mu\nu} \gamma_5 q_B^b) \gamma^\nu \gamma_5 c^c \, ,
\\
B^U_{\mathbf{6},7\mu\nu} &=& \epsilon_{abc} S^U_{AB} (q_A^{aT} \mathbb{C} \sigma_{\mu\nu} q_B^b) \gamma_5 c^c \, ,
\\
 B^U_{\mathbf{6},8\mu\nu} &=& \epsilon_{abc} S^U_{AB} (q_A^{aT} \mathbb{C} \sigma_{\mu\nu} \gamma_5 q_B^b) c^c \, .
\end{eqnarray}
In the above expressions, $a,b,c$ are color indices and the sum over repeated indices is taken;
$A,B,G,U$ are $SU(3)$ flavor indices, so that $q_A = \{u,d,s\}$;
$\epsilon^{ABG}$ is the totally antisymmetric matrix with $G=1,2,3$, so that $B^G_{\mathbf{\bar 3},i}$ belong to the $SU(3)$ flavor $\mathbf{\bar 3}_F$ representation;
$S^U_{AB}$ are the totally symmetric matrices with $U=1\cdots6$, so that $B^U_{\mathbf{6},i}$ belong to the $SU(3)$ flavor $\mathbf{6}_F$ representation;
$c^c$ is the charm quark field with the color index $c$;
$\mathbb{C}$ is the charge-conjugation matrix; $P^{3/2}_{\mu\nu}$ and $P^{3/2}_{\mu\nu\alpha\beta}$ are two $J=3/2$ projection operators.

Among the nine fields given in Eqs.~(\ref{def:heavy1}-\ref{def:heavy9}), 
$B^G_{\mathbf{\bar 3},1}$, $B^G_{\mathbf{\bar 3},2}$, $B^G_{\mathbf{\bar 3},3}$, $B^U_{\mathbf{6},4}$, and $B^U_{\mathbf{6},5}$ have pure spin $J=1/2$, and $B^G_{\mathbf{\bar 3},\mu}$, $B^U_{\mathbf{6},\mu}$, $B^{\prime U}_{\mathbf{6},\mu}$, and $B^U_{\mathbf{6},\mu\nu}$ have pure spin $J=3/2$.
In the present study we only take into account the $J^P = 1/2^+$ ``ground-state'' charmed baryon fields, $B^G_{\mathbf{\bar 3},2}$ and $B^U_{\mathbf{6},4}$; while we omit other charmed baryon fields, $B^G_{\mathbf{\bar 3},1}$, $B^G_{\mathbf{\bar 3},3}$, $B^G_{\mathbf{\bar 3},\mu}$, $B^U_{\mathbf{6},5}$, $B^{U}_{\mathbf{6},\mu}$, $B^{\prime U}_{\mathbf{6},\mu}$, and $B^U_{\mathbf{6},\mu\nu}$, all of which do not couple to the $J^P = 1/2^+$ ground-state charmed baryons $\Lambda_c$ and $\Sigma_c$ within the framework of heavy quark effective theory~\cite{groundbaryon}.

Then we give the relations among light baryon fields. We refer to Refs.~\cite{Ioffe:1981kw,Ioffe:1982ce,Espriu:1983hu,Chen:2008qv,Chen:2009sf,Chen:2010ba,Chen:2011rh,Dmitrasinovic:2016hup} for detailed discussions. According to the results of Ref.~\cite{Chen:2008qv}, we can use $u$, $u$, and $d$ ($q=u/d$) quarks to construct five independent baryon fields:
\begin{eqnarray}
N_1 &=& \epsilon^{abc} (u_a^T \mathbb{C} d_b) \gamma_5 u_c \, ,
\\ N_2 &=& \epsilon^{abc} (u_a^T \mathbb{C} \gamma_5 d_b) u_c \, ,
\\ N_{3}^{\prime\mu} &=& \epsilon^{abc} (u_a^T \mathbb{C} \gamma^\mu \gamma_5 d_b) \gamma_5 u_c \, ,
\\ N_{4}^{\prime\mu} &=& \epsilon^{abc} (u_a^T \mathbb{C} \gamma^\mu d_b) u_c \, ,
\\ N_{5}^{\prime\mu\nu} &=& \epsilon^{abc} (u_a^T \mathbb{C} \sigma^{\mu\nu} d_b) \gamma_5 u_c \, .
\end{eqnarray}
Among these fields, the former two $N_{1,2}$ have pure spin $J=1/2$, but the latter three $N^{\prime\mu(\nu)}_{3,4,5}$ do not have pure spin $J=3/2$. We need to further use the projection operators $P_{3/2}^{\mu\alpha}$ and $P_{3/2}^{\mu\nu\alpha\beta}$ to obtain $N^{\mu(\nu)}_{3,4,5}$, already given in Eqs.~(\ref{eq:lightbaryon}), which have pure spin $J=3/2$. The relations between $N^{\mu(\nu)}_{3,4,5}$ and $N^{\prime\mu(\nu)}_{3,4,5}$ are
\begin{eqnarray}
N_3^\mu &=& P_{3/2}^{\mu\alpha} \times N_{3\alpha}^{\prime}
\\ \nonumber &=& N_{3}^{\prime\mu} + {1\over4} \gamma^\mu \gamma_5 (N_1 - N_2) \, ,
\\ N_4^\mu &=& P_{3/2}^{\mu\alpha} \times N_{4\alpha}^{\prime}
\\ \nonumber &=& N_{4}^{\prime\mu} + {1\over4} \gamma^\mu \gamma_5 (N_1 - N_2) \, ,
\\ N_5^{\mu\nu} &=& P_{3/2}^{\mu\nu\alpha\beta} \times N_{5\alpha\beta}^{\prime}
\\ \nonumber &=& N_{5}^{\prime\mu\nu} + {i\over2} \gamma^\nu \gamma_5 (N_{3}^{\prime\mu} + N_{4}^{\prime\mu})
\\ \nonumber && - {i\over2} \gamma^\mu \gamma_5 (N_{3}^{\prime\nu} + N_{4}^{\prime\nu}) + {1\over3} \sigma^{\mu\nu} (2 N_1 - N_2) \, .
\end{eqnarray}
All the other baryon fields can be transformed to $N^{(\prime\mu\nu)}_{1,2,3,4,5}$ \Big(and so to $N^{(\mu\nu)}_{1,2,3,4,5}$\Big) through:
\begin{eqnarray}
&& \epsilon^{abc} (u_a^T \mathbb{C} \gamma^\mu d_b) \gamma_\mu \gamma_5 u_c = N_1 - N_2 \, ,
\label{eq:lightbaryon1}
\\ \nonumber && \epsilon^{abc} (u_a^T \mathbb{C} \gamma^\mu \gamma_5 d_b) \gamma_\mu u_c = N_1 - N_2 \, ,
\\ \nonumber && \epsilon^{abc} (u_a^T \mathbb{C} \sigma^{\mu\nu} d_b) \sigma_{\mu\nu} \gamma_5 u_c = - 2 N_1 - 2 N_2 \, ,
\\ \nonumber && \epsilon^{abc} (u_a^T \mathbb{C} \gamma_\nu d_b) \sigma^{\mu\nu} u_c
\\ \nonumber && ~~~~~~~~~~ = - i N_{4}^{\prime\mu} - i \gamma^\mu \gamma_5 (N_1 - N_2) \, ,
\\ \nonumber && \epsilon^{abc} (u_a^T \mathbb{C} \gamma_\nu \gamma_5 d_b) \sigma^{\mu\nu} \gamma_5 u_c
\\ \nonumber && ~~~~~~~~~~ = - i N_{3}^{\prime\mu} - i \gamma^\mu \gamma_5 (N_1 - N_2) \, ,
\\ \nonumber && \epsilon^{abc} (u_a^T \mathbb{C} \sigma^{\mu\nu} d_b) \gamma_\nu u_c
\\ \nonumber && ~~~~~~~~~~ = i N_{3}^{\prime\mu} + i N_{4}^{\prime\mu} + i \gamma^\mu \gamma_5 N_1 \, ,
\\ \nonumber && \epsilon^{abc} (u_a^T \mathbb{C} \sigma^{\mu\nu} \gamma_5 d_b) \gamma_\nu \gamma_5 u_c
\\ \nonumber && ~~~~~~~~~~ = - i N_{3}^{\prime\mu} - i N_{4}^{\prime\mu} + i \gamma^\mu \gamma_5 N_2 \, ,
\\ \nonumber && \epsilon^{abc} (u_a^T \mathbb{C} \sigma^{\mu\nu} \gamma_5 d_b) u_c
\\ \nonumber && ~~~~~~~~~~ = N_{5}^{\prime\mu\nu} + i \gamma^\nu \gamma_5 (N_{3}^{\prime\mu} + N_{4}^{\prime\mu})
\\ \nonumber && ~~~~~~~~~~~~~ - i \gamma^\mu \gamma_5 (N_{3}^{\prime\nu} + N_{4}^{\prime\nu}) + \sigma^{\mu\nu} (N_1 - N_2) \, ,
\\ \nonumber && \epsilon^{abc} \epsilon^{\mu\nu\rho\sigma} (u_a^T \mathbb{C} \sigma_{\rho\alpha} d_b) \sigma_{\sigma\alpha} u_c
\\ \nonumber && ~~~~~~~~~~ = - 2 N_{5}^{\prime\mu\nu} - i \gamma^\nu \gamma_5 (N_{3}^{\prime\mu} + N_{4}^{\prime\mu})
\\ \nonumber && ~~~~~~~~~~~~~ + i \gamma^\mu \gamma_5 (N_{3}^{\prime\nu} + N_{4}^{\prime\nu}) - 2 \sigma^{\mu\nu} N_1 \, ,
\\ \nonumber && \epsilon^{abc} \epsilon^{\mu\nu\rho\sigma} (u_a^T \mathbb{C} \gamma_\rho d_b) \gamma_\sigma u_c
\\ \nonumber && ~~~~~~~~~~ = - i \gamma^\nu \gamma_5 N_{4}^{\prime\mu} + i \gamma^\mu \gamma_5 N_{4}^{\prime\nu} - \sigma^{\mu\nu} (N_1 - N_2) \, ,
\\ \nonumber && \epsilon^{abc} \epsilon^{\mu\nu\rho\sigma} (u_a^T \mathbb{C} \gamma_\rho \gamma_5 d_b) \gamma_\sigma \gamma_5 u_c
\\ \nonumber && ~~~~~~~~~~ = - i \gamma^\nu \gamma_5 N_{3}^{\prime\mu} + i \gamma^\mu \gamma_5 N_{3}^{\prime\nu} - \sigma^{\mu\nu} (N_1 - N_2) \, ,
\end{eqnarray}
and
\begin{eqnarray}
&& \epsilon^{abc} (u_a^T \mathbb{C} u_b) \gamma_5 d_c = 0 \, ,
\label{eq:lightbaryon2}
\\ \nonumber && \epsilon^{abc} (u_a^T \mathbb{C} \gamma_5 u_b) d_c = 0 \, ,
\\ \nonumber && \epsilon^{abc} (u_a^T \mathbb{C} \gamma^\mu u_b) \gamma_\mu \gamma_5 d_c = - 2 N_1 + 2 N_2 \, ,
\\ \nonumber && \epsilon^{abc} (u_a^T \mathbb{C} \gamma^\mu \gamma_5 u_b) \gamma_\mu d_c = 0 \, ,
\\ \nonumber && \epsilon^{abc} (u_a^T \mathbb{C} \sigma^{\mu\nu} u_b) \sigma_{\mu\nu} \gamma_5 d_c = 4 N_1 + 4 N_2 \, ,
\\ \nonumber && \epsilon^{abc} (u_a^T \mathbb{C} \gamma^\mu u_b) d_c
\\ \nonumber && ~~~~~~~~~~ =  N_{3}^{\prime\mu} + N_{4}^{\prime\mu} + \gamma^\mu \gamma_5 (N_1 - N_2) \, ,
\\ \nonumber && \epsilon^{abc} (u_a^T \mathbb{C} \gamma^\mu \gamma_5 u_b) \gamma_5 d_c = 0 \, ,
\\ \nonumber && \epsilon^{abc} (u_a^T \mathbb{C} \gamma_\nu u_b) \sigma^{\mu\nu} d_c
\\ \nonumber && ~~~~~~~~~~ =  - i N_{3}^{\prime\mu} - i N_{4}^{\prime\mu} + i \gamma^\mu \gamma_5 (N_1 - N_2) \, ,
\\ \nonumber && \epsilon^{abc} (u_a^T \mathbb{C} \gamma_\nu \gamma_5 u_b) \sigma^{\mu\nu} \gamma_5 d_c = 0 \, ,
\\ \nonumber && \epsilon^{abc} (u_a^T \mathbb{C} \sigma^{\mu\nu} u_b) \gamma_\nu d_c
\\ \nonumber && ~~~~~~~~~~ =  - i N_{3}^{\prime\mu} + i N_{4}^{\prime\mu} - i \gamma^\mu \gamma_5 (N_1 + N_2) \, ,
\\ \nonumber && \epsilon^{abc} (u_a^T \mathbb{C} \sigma^{\mu\nu} \gamma_5 u_b) \gamma_\nu \gamma_5 d_c
\\ \nonumber && ~~~~~~~~~~ =  i N_{3}^{\prime\mu} - i N_{4}^{\prime\mu} - i \gamma^\mu \gamma_5 (N_1 + N_2) \, ,
\\ \nonumber && \epsilon^{abc} (u_a^T \mathbb{C} \sigma^{\mu\nu} u_b) \gamma_5 d_c
\\ \nonumber && ~~~~~~~~~~ =  N_{5}^{\prime\mu\nu} + i \gamma^\nu \gamma_5 N_{3}^{\prime\mu} - i \gamma^\mu \gamma_5 N_{3}^{\prime\nu} + \sigma^{\mu\nu} N_1 \, ,
\\ \nonumber && \epsilon^{abc} (u_a^T \mathbb{C} \sigma^{\mu\nu} \gamma_5 u_b) d_c
\\ \nonumber && ~~~~~~~~~~ =  N_{5}^{\prime\mu\nu} + i \gamma^\nu \gamma_5 N_{4}^{\prime\mu} - i \gamma^\mu \gamma_5 N_{4}^{\prime\nu} + \sigma^{\mu\nu} N_1 \, ,
\\ \nonumber && \epsilon^{abc} \epsilon^{\mu\nu\rho\sigma} (u_a^T \mathbb{C} \sigma_{\rho\alpha} u_b) \sigma_{\sigma\alpha} d_c
\\ \nonumber && ~~~~~~~~~~ = - 2 N_{5}^{\prime\mu\nu} - i \gamma^\nu \gamma_5 (N_{3}^{\prime\mu} + N_{4}^{\prime\mu})
\\ \nonumber && ~~~~~~~~~~~~~  + i \gamma^\mu \gamma_5 (N_{3}^{\prime\nu} + N_{4}^{\prime\nu}) + 2 \sigma^{\mu\nu} N_2 \, ,
\\ \nonumber && \epsilon^{abc} \epsilon^{\mu\nu\rho\sigma} (u_a^T \mathbb{C} \gamma_\rho u_b) \gamma_\sigma d_c
\\ \nonumber && ~~~~~~~~~~ =  - i \gamma^\nu \gamma_5 (N_{3}^{\prime\mu} + N_{4}^{\prime\mu}) + i \gamma^\mu \gamma_5 (N_{3}^{\prime\nu} + N_{4}^{\prime\nu}) \, ,
\\ \nonumber && \epsilon^{abc} \epsilon^{\mu\nu\rho\sigma} (u_a^T \mathbb{C} \gamma_\rho \gamma_5 u_b) \gamma_\sigma \gamma_5 d_c = 0 \, .
\end{eqnarray}

\section{Uncertainties due to phase angles}
\label{app:phase}

There are two different effective Lagrangians for the $|\bar D^0 \Sigma_c^{+}; 1/2^- \rangle$ (and $|\bar D^- \Sigma_c^{++}; 1/2^- \rangle$) decay into the $\eta_c p$ final state, as given in Eqs.~(\ref{lag:etacpA}) and (\ref{lag:etacpB}):
\begin{eqnarray}
\mathcal{L}_{\eta_c p} &=& g_{\eta_c p}~\bar P_c N~\eta_c \, ,
\\ \mathcal{L}^\prime_{\eta_c p} &=& g^\prime_{\eta_c p}~\bar P_c \gamma_\mu N~\partial^\mu\eta_c \, .
\end{eqnarray}
There are also two different effective Lagrangians for the $|\bar D^{*0} \Sigma_c^{+}; 1/2^- \rangle$ (and $|\bar D^{*-} \Sigma_c^{++}; 1/2^- \rangle$) decay into the $J/\psi p$ final state, as given in Eqs.~(\ref{lag:psipA}) and (\ref{lag:psipB}):
\begin{eqnarray}
\mathcal{L}_{\psi p} &=& g_{\psi p}~\bar P_c \gamma_\mu \gamma_5 N~\psi^\mu \, ,
\\ \mathcal{L}^\prime_{\psi p} &=& g^\prime_{\psi p}~\bar P_c \sigma_{\mu\nu} \gamma_5 N~\partial^\mu\psi^\nu \, .
\end{eqnarray}
There can be a phase angle $\theta$ between $g_{\eta_c p}$ and $g^\prime_{\eta_c p}$ and another phase angle $\theta^\prime$ between $g_{\psi p}$ and $g^\prime_{\psi p}$, both of which can not be determined in the present study. In this appendix we rotate $\theta/\theta^\prime$ and redo all the calculations.

\begin{widetext}
\begin{itemize}

\item We obtain the following relative branching ratios for the $\bar D^{(*)} \Sigma_c$ hadronic molecular states of $I = 1/2$:
\begin{eqnarray}
&& {\mathcal{B}\left(|\bar D \Sigma_c; 1/2^- \rangle \rightarrow
~~~J/\psi p~~~
: ~~~~~~~~\eta_c p~~~~~~~~
: ~~~\bar D^{*0} \Lambda_c^+~
\right) \over \mathcal{B}\left(|\bar D \Sigma_c; 1/2^- \rangle \rightarrow J/\psi p\right)}
\\[2mm] \nonumber &\approx&
~~~~~~~~~~~~~~~~~~~~~~~~~~~~~~~ 1 ~~~~~ : ~~~~~0.5\sim3.8~~~~ : \,~~~~0.69t~ \, ,
\\[2mm]
&& {\mathcal{B}\left(|\bar D^* \Sigma_c; 1/2^- \rangle \rightarrow
~~J/\psi p~~
: \,~~~~\eta_c p~~~~\,
: \,~\chi_{c0} p~\,
: ~\bar D^{0} \Lambda_c^+~
: ~\bar D^{*0} \Lambda_c^+~
: ~\bar D^{0} \Sigma_c^+~
: ~\bar D^{-} \Sigma_c^{++}~
\right) \over \mathcal{B}\left(|\bar D^* \Sigma_c; 1/2^- \rangle \rightarrow J/\psi p\right)\big|_{\theta^\prime = 0}}
\\[2mm] \nonumber &\approx&
~~~~~~~~~~~~~~~~~~~~~~~~~~ 1\sim1.8 ~\, : ~0.1\sim1.1~ : ~0.004~ : \,~~1.2t~~\, : \,~~0.41t~~\, : ~~0.04t~~ : \,~~~0.08t~ \, ,
\\[2mm]
&& {\mathcal{B}\left(|\bar D^* \Sigma_c; 3/2^- \rangle \rightarrow
~~J/\psi p~~
: ~~\eta_c p~~
: \,~\chi_{c1} p~\,
: ~\bar D^{*0} \Lambda_c^+~
: ~\bar D^{0} \Sigma_c^+~
: ~\bar D^{-} \Sigma_c^{++}~
\right) \over \mathcal{B}\left(|\bar D^* \Sigma_c; 3/2^- \rangle \rightarrow J/\psi p\right)}
\\[2mm] \nonumber &\approx&
~~~~~~~~~~~~~~~~~~~~~~~~~~~~~~~ 1 ~~~~\, : ~0.005~ : ~10^{-4}~ : \,~~0.35t~~\, : \,~10^{-5}t~\, : ~~~10^{-5}t~ \, .
\end{eqnarray}

\item We obtain the following relative branching ratios for the $\bar D^{(*)} \Sigma_c$ hadronic molecular states of $I = 3/2$:
\begin{eqnarray}
{\mathcal{B}\left(|\bar D^* \Sigma_c; 1/2^{-\prime} \rangle \rightarrow \bar D^{-} \Sigma_c^{++} \right) \over \mathcal{B}\left(|\bar D^* \Sigma_c; 1/2^{-\prime} \rangle \rightarrow \bar D^{0} \Sigma_c^+ \right)} &\approx& 0.5 \, ,
\\[2mm]
{\mathcal{B}\left(|\bar D^* \Sigma_c; 3/2^{-\prime} \rangle \rightarrow \bar D^{-} \Sigma_c^{++} \right) \over \mathcal{B}\left(|\bar D^* \Sigma_c; 3/2^{-\prime} \rangle \rightarrow \bar D^{0} \Sigma_c^+ \right)} &\approx& 0.5 \, .
\end{eqnarray}

\item We obtain the following relative branching ratios for the $\bar D^{(*)0} \Sigma_c^+$ hadronic molecular states:
\begin{eqnarray}
&& {\mathcal{B}\left(|\bar D^0 \Sigma_c^+; 1/2^- \rangle \rightarrow
~~~J/\psi p~~~
: ~~~~~~~~\eta_c p~~~~~~~~
: ~~~\bar D^{*0} \Lambda_c^+~
\right) \over \mathcal{B}\left(|\bar D^0 \Sigma_c^+; 1/2^- \rangle \rightarrow J/\psi p\right)}
\\[2mm] \nonumber &\approx&
~~~~~~~~~~~~~~~~~~~~~~~~~~~~~~~~~ 1 ~~~~~ : ~~~~~0.5\sim3.8~~~~ : \,~~~~0.69t~ \, ,
\\[2mm]
&& {\mathcal{B}\left(|\bar D^{*0} \Sigma_c^+; 1/2^- \rangle \rightarrow
~~J/\psi p~~
: \,~~~~\eta_c p~~~~\,
: \,~\chi_{c0} p~\,
: ~\bar D^{0} \Lambda_c^+~
: ~\bar D^{*0} \Lambda_c^+~
: ~\bar D^{0} \Sigma_c^+~
: ~\bar D^{-} \Sigma_c^{++}~
\right) \over \mathcal{B}\left(|\bar D^{*0} \Sigma_c^+; 1/2^- \rangle \rightarrow J/\psi p\right)\big|_{\theta^\prime = 0}}
\\[2mm] \nonumber &\approx&
~~~~~~~~~~~~~~~~~~~~~~~~~~~~ 1\sim1.8 ~\, : ~0.1\sim1.1~ : ~0.004~ : \,~~1.2t~~\, : \,~~0.41t~~\, : ~~0.35t~~ : \,~~~0.70t~ \, ,
\\[2mm]
&& {\mathcal{B}\left(|\bar D^{*0} \Sigma_c^+; 3/2^- \rangle \rightarrow
~~J/\psi p~~
: ~~\eta_c p~~
: \,~\chi_{c1} p~\,
: ~\bar D^{*0} \Lambda_c^+~
: ~\bar D^{0} \Sigma_c^+~
: ~\bar D^{-} \Sigma_c^{++}~
\right) \over \mathcal{B}\left(|\bar D^{*0} \Sigma_c^+; 3/2^- \rangle \rightarrow J/\psi p\right)}
\\[2mm] \nonumber &\approx&
~~~~~~~~~~~~~~~~~~~~~~~~~~~~~~~~~ 1 ~~~~\, : ~0.005~ : ~10^{-4}~ : \,~~0.35t~~\, : \,~10^{-4}t~\, : ~~~10^{-4}t~ \, .
\end{eqnarray}

\item We obtain the following relative branching ratios for the $\bar D^{(*)-} \Sigma_c^{++}$ hadronic molecular states:
\begin{eqnarray}
&& {\mathcal{B}\left(|\bar D^- \Sigma_c^{++}; 1/2^- \rangle \rightarrow
~~~J/\psi p~~~
: ~~~~~~~~\eta_c p~~~~~~~~
: ~~~\bar D^{*0} \Lambda_c^+~
\right) \over \mathcal{B}\left(|\bar D^- \Sigma_c^{++}; 1/2^- \rangle \rightarrow J/\psi p\right)}
\\[2mm] \nonumber &\approx&
~~~~~~~~~~~~~~~~~~~~~~~~~~~~~~~~~~~\, 1 ~~~~~ : ~~~~~0.5\sim3.8~~~~ : \,~~~~0.69t~ \, ,
\\[2mm]
&& {\mathcal{B}\left(|\bar D^{*-} \Sigma_c^{++}; 1/2^- \rangle \rightarrow
~~J/\psi p~~
: \,~~~~\eta_c p~~~~\,
: \,~\chi_{c0} p~\,
: ~\bar D^{0} \Lambda_c^+~
: ~\bar D^{*0} \Lambda_c^+~
: ~\bar D^{0} \Sigma_c^+~
: ~\bar D^{-} \Sigma_c^{++}~
\right) \over \mathcal{B}\left(|\bar D^{*-} \Sigma_c^{++}; 1/2^- \rangle \rightarrow J/\psi p\right)\big|_{\theta^\prime = 0}}
\\[2mm] \nonumber &\approx&
~~~~~~~~~~~~~~~~~~~~~~~~~~~~~~\, 1\sim1.8 ~\, : ~0.1\sim1.1~ : ~0.004~ : \,~~1.2t~~\, : \,~~0.41t~~\, : ~~0.35t~~ : \,~~~~~0~ \, ,
\\[2mm]
&& {\mathcal{B}\left(|\bar D^{*-} \Sigma_c^{++}; 3/2^- \rangle \rightarrow
~~J/\psi p~~
: ~~\eta_c p~~
: \,~\chi_{c1} p~\,
: ~\bar D^{*0} \Lambda_c^+~
: ~\bar D^{0} \Sigma_c^+~
: ~\bar D^{-} \Sigma_c^{++}~
\right) \over \mathcal{B}\left(|\bar D^{*-} \Sigma_c^{++}; 3/2^- \rangle \rightarrow J/\psi p\right)}
\\[2mm] \nonumber &\approx&
~~~~~~~~~~~~~~~~~~~~~~~~~~~~~~~~~~~\, 1 ~~~~\, : ~0.005~ : ~10^{-4}~ : \,~~0.35t~~\, : \,~10^{-4}t~\, : ~~~~~0~ \, .
\end{eqnarray}

\end{itemize}
\end{widetext}


\begin{thebibliography}{99}

\bibitem{Choi:2003ue}
  S.~K.~Choi {\it et al.} [Belle Collaboration],
  {Phys.\ Rev.\ Lett.\  {\bf 91}, 262001 (2003)}.

\bibitem{pdg}
  M.~Tanabashi {\it et al.} [Particle Data Group],
  {Phys.\ Rev.\ D {\bf 98}, 030001 (2018)}.

\bibitem{Aaij:2015tga}
  R.~Aaij {\it et al.} [LHCb Collaboration],
  {Phys.\ Rev.\ Lett.\  {\bf 115}, 072001 (2015)}.

\bibitem{Aaij:2019vzc}
  R.~Aaij {\it et al.} [LHCb Collaboration],
  {Phys.\ Rev.\ Lett.\  {\bf 122}, 222001 (2019)}.

\bibitem{Chen:2016qju}
  H.~X.~Chen, W.~Chen, X.~Liu and S.~L.~Zhu,
  {Phys.\ Rept.\  {\bf 639}, 1 (2016)}.

\bibitem{Liu:2019zoy}
  Y.~R.~Liu, H.~X.~Chen, W.~Chen, X.~Liu and S.~L.~Zhu,
  {Prog.\ Part.\ Nucl.\ Phys.\  {\bf 107}, 237 (2019)}.

\bibitem{Lebed:2016hpi}
  R.~F.~Lebed, R.~E.~Mitchell and E.~S.~Swanson,
  {Prog.\ Part.\ Nucl.\ Phys.\  {\bf 93}, 143 (2017)}.

\bibitem{Esposito:2016noz}
  A.~Esposito, A.~Pilloni and A.~D.~Polosa,
  {Phys.\ Rept.\  {\bf 668}, 1 (2017)}.

\bibitem{Guo:2017jvc}
  F.~K.~Guo, C.~Hanhart, U.~G.~Mei{\ss}ner, Q.~Wang, Q.~Zhao, and B.~S.~Zou,
  {Rev.\ Mod.\ Phys.\  {\bf 90}, 015004 (2018)}.

\bibitem{Ali:2017jda}
  A.~Ali, J.~S.~Lange and S.~Stone,
  {Prog.\ Part.\ Nucl.\ Phys.\  {\bf 97}, 123 (2017)}.

\bibitem{Olsen:2017bmm}
  S.~L.~Olsen, T.~Skwarnicki, and D.~Zieminska,
  {Rev.\ Mod.\ Phys.\  {\bf 90}, 015003 (2018)}.

\bibitem{Karliner:2017qhf}
  M.~Karliner, J.~L.~Rosner and T.~Skwarnicki,
  {Ann.\ Rev.\ Nucl.\ Part.\ Sci.\  {\bf 68}, 17 (2018)}.

\bibitem{Brambilla:2019esw}
  N.~Brambilla, S.~Eidelman, C.~Hanhart, A.~Nefediev, C.~P.~Shen, C.~E.~Thomas, A.~Vairo and C.~Z.~Yuan,
  {arXiv:1907.07583 [hep-ex]}.

\bibitem{Guo:2019twa}
  F.~K.~Guo, X.~H.~Liu and S.~Sakai,
  {arXiv:1912.07030 [hep-ph]}.

\bibitem{Wu:2010jy}
  J.~J.~Wu, R.~Molina, E.~Oset and B.~S.~Zou,
  {Phys.\ Rev.\ Lett.\  {\bf 105}, 232001 (2010)}.

\bibitem{Wang:2011rga}
  W.~L.~Wang, F.~Huang, Z.~Y.~Zhang and B.~S.~Zou,
  {Phys.\ Rev.\ C {\bf 84}, 015203 (2011)}.

\bibitem{Yang:2011wz}
  Z.~C.~Yang, Z.~F.~Sun, J.~He, X.~Liu and S.~L.~Zhu,
  {Chin.\ Phys.\ C {\bf 36}, 6 (2012)}.

\bibitem{Karliner:2015ina}
  M.~Karliner and J.~L.~Rosner,
  {Phys.\ Rev.\ Lett.\  {\bf 115}, 122001 (2015)}.

\bibitem{Wu:2012md}
  J.~J.~Wu, T.-S.~H.~Lee and B.~S.~Zou,
  {Phys.\ Rev.\ C {\bf 85}, 044002 (2012)}.

\bibitem{Chen:2019asm}
  R.~Chen, Z.~F.~Sun, X.~Liu and S.~L.~Zhu,
  {Phys.\ Rev.\ D {\bf 100}, 011502(R) (2019)}.

\bibitem{Liu:2019tjn}
  M.~Z.~Liu, Y.~W.~Pan, F.~Z.~Peng, M.~S.~S{\'a}nchez, L.~S.~Geng, A.~Hosaka and M.~P.~Valderrama,
  {Phys.\ Rev.\ Lett.\  {\bf 122}, 242001 (2019)}.

\bibitem{He:2019ify}
  J.~He,
  {Eur.\ Phys.\ J.\ C {\bf 79}, 393 (2019)}.

\bibitem{Huang:2019jlf}
  H.~Huang, J.~He and J.~Ping,
  {arXiv:1904.00221 [hep-ph]}.

\bibitem{Guo:2019kdc}
  Z.~H.~Guo and J.~A.~Oller,
  {Phys.\ Lett.\ B {\bf 793}, 144 (2019)}.

\bibitem{Fernandez-Ramirez:2019koa}
  C.~Fern\'andez-Ram\'{\i}rez, A.~Pilloni, M.~Albaladejo, A.~Jackura, V.~Mathieu, M.~Mikhasenko, J.~A.~Silva-Castro, A.~P.~Szczepaniak, [JPAC Collaboration],
  {Phys.\ Rev.\ Lett.\  {\bf 123}, 092001 (2019)}.

\bibitem{Xiao:2019aya}
  C.~W.~Xiao, J.~Nieves and E.~Oset,
  {Phys.\ Rev.\ D {\bf 100}, 014021 (2019)}.

\bibitem{Meng:2019ilv}
  L.~Meng, B.~Wang, G.~J.~Wang and S.~L.~Zhu,
  {Phys.\ Rev.\ D {\bf 100}, 014031 (2019)}.

\bibitem{Wu:2019adv}
  J.~J.~Wu, T.-S.~H.~Lee and B.~S.~Zou,
  {Phys.\ Rev.\ C {\bf 100}, 035206 (2019)}.

\bibitem{Wang:2019hyc}
  Z.~G.~Wang and X.~Wang,
  {arXiv:1907.04582 [hep-ph]}.

\bibitem{Yamaguchi:2019seo}
  Y.~Yamaguchi, H.~Garcia-Tecocoatzi, A.~Giachino, A.~Hosaka, E.~Santopinto, S.~Takeuchi and M.~Takizawa,
  {arXiv:1907.04684 [hep-ph]}.

\bibitem{Valderrama:2019chc}
  M.~Pavon Valderrama,
  {Phys.\ Rev.\ D {\bf 100}, 094028 (2019)}.

\bibitem{Liu:2019zvb}
  M.~Z.~Liu, T.~W.~Wu, M.~S{\'a}nchez S{\'a}nchez, M.~P.~Valderrama, L.~S.~Geng and J.~J.~Xie,
  {arXiv:1907.06093 [hep-ph]}.

\bibitem{Burns:2019iih}
  T.~J.~Burns and E.~S.~Swanson,
  {Phys.\ Rev.\ D {\bf 100}, 114033 (2019)}.

\bibitem{Wang:2019ato}
  B.~Wang, L.~Meng and S.~L.~Zhu,
  {JHEP {\bf 1911}, 108 (2019)}.

\bibitem{Gutsche:2019mkg}
  T.~Gutsche and V.~E.~Lyubovitskij,
  {Phys.\ Rev.\ D {\bf 100}, 094031 (2019)}.

\bibitem{Du:2019pij}
  M.~L.~Du, V.~Baru, F.~K.~Guo, C.~Hanhart, U.~G.~Mei{\ss}ner, J.~A.~Oller and Q.~Wang,
  {arXiv:1910.11846 [hep-ph]}.

\bibitem{Azizi:2016dhy}
  K.~Azizi, Y.~Sarac and H.~Sundu,
  {Phys.\ Rev.\ D {\bf 95}, 094016 (2017)}.

\bibitem{Chen:2019bip}
  H.~X.~Chen, W.~Chen and S.~L.~Zhu,
  {Phys.\ Rev.\ D {\bf 100}, 051501(R) (2019)}.

\bibitem{Maiani:2015vwa}
  L.~Maiani, A.~D.~Polosa and V.~Riquer,
  {Phys.\ Lett.\ B {\bf 749}, 289 (2015)}.

\bibitem{Lebed:2015tna}
  R.~F.~Lebed,
  {Phys.\ Lett.\ B {\bf 749}, 454 (2015)}.

\bibitem{Stancu:2019qga}
  F.~Stancu,
  {Eur.\ Phys.\ J.\ C {\bf 79}, 957 (2019)}.

\bibitem{Giron:2019bcs}
  J.~F.~Giron, R.~F.~Lebed and C.~T.~Peterson,
  {JHEP {\bf 1905}, 061 (2019)}.

\bibitem{Ali:2019npk}
  A.~Ali and A.~Y.~Parkhomenko,
  {Phys.\ Lett.\ B {\bf 793}, 365 (2019)}.

\bibitem{Weng:2019ynv}
  X.~Z.~Weng, X.~L.~Chen, W.~Z.~Deng and S.~L.~Zhu,
  {Phys.\ Rev.\ D {\bf 100}, 016014 (2019)}.

\bibitem{Eides:2019tgv}
  M.~I.~Eides, V.~Y.~Petrov and M.~V.~Polyakov,
  {arXiv:1904.11616 [hep-ph]}.

\bibitem{Wang:2019got}
  Z.~G.~Wang,
  {arXiv:1905.02892 [hep-ph]}.

\bibitem{Cheng:2019obk}
  J.~B.~Cheng and Y.~R.~Liu,
  {Phys.\ Rev.\ D {\bf 100}, 054002 (2019)}.

\bibitem{Ali:2019clg}
  A.~Ali, I.~Ahmed, M.~J.~Aslam, A.~Y.~Parkhomenko and A.~Rehman,
  {JHEP {\bf 1910}, 256 (2019)}.

\bibitem{Voloshin:2019aut}
  M.~B.~Voloshin,
  {Phys.\ Rev.\ D {\bf 100}, 034020 (2019)}.

\bibitem{Sakai:2019qph}
  S.~Sakai, H.~J.~Jing and F.~K.~Guo,
  {Phys.\ Rev.\ D {\bf 100}, 074007 (2019)}.

\bibitem{Guo:2019fdo}
  F.~K.~Guo, H.~J.~Jing, U.~G.~Mei{\ss}ner and S.~Sakai,
  {Phys.\ Rev.\ D {\bf 99}, 091501(R) (2019)}.

\bibitem{Xiao:2019mst}
  C.~J.~Xiao, Y.~Huang, Y.~B.~Dong, L.~S.~Geng and D.~Y.~Chen,
  {Phys.\ Rev.\ D {\bf 100}, 014022 (2019)}.

\bibitem{Cao:2019kst}
  X.~Cao and J.~P.~Dai,
  {Phys.\ Rev.\ D {\bf 100}, 054033 (2019)}.

\bibitem{Lin:2019qiv}
  Y.~H.~Lin and B.~S.~Zou,
  {Phys.\ Rev.\ D {\bf 100}, 056005 (2019)}.

\bibitem{Xu:2019zme}
  Y.~J.~Xu, C.~Y.~Cui, Y.~L.~Liu and M.~Q.~Huang,
  {arXiv:1907.05097 [hep-ph]}.

\bibitem{Wang:2019spc}
  G.~J.~Wang, L.~Y.~Xiao, R.~Chen, X.~H.~Liu, X.~Liu and S.~L.~Zhu,
  {arXiv:1911.09613 [hep-ph]}.

\bibitem{Chen:2019wjd}
  H.~X.~Chen,
  {arXiv:1910.03269 [hep-ph]}.

\bibitem{Voloshin:2013dpa}
  M.~B.~Voloshin,
  {Phys.\ Rev.\ D {\bf 87}, 091501(R) (2013)}.

\bibitem{Maiani:2017kyi}
  L.~Maiani, A.~D.~Polosa and V.~Riquer,
  {Phys.\ Lett.\ B {\bf 778}, 247 (2018)}.

\bibitem{Voloshin:2018pqn}
  M.~B.~Voloshin,
  {Phys.\ Rev.\ D {\bf 98}, 034025 (2018)}.

\bibitem{Xiao:2019spy}
  L.~Y.~Xiao, G.~J.~Wang and S.~L.~Zhu,
  {arXiv:1912.12781 [hep-ph]}.

\bibitem{Cheng:2020nho}
  J.~B.~Cheng, S.~Y.~Li, Y.~R.~Liu, Y.~N.~Liu, Z.~G.~Si and T.~Yao,
  {arXiv:2001.05287 [hep-ph]}.

\bibitem{Chen:2015moa}
  H.~X.~Chen, W.~Chen, X.~Liu, T.~G.~Steele and S.~L.~Zhu,
  {Phys.\ Rev.\ Lett.\  {\bf 115}, 172001 (2015)}.

\bibitem{Chen:2016otp}
  H.~X.~Chen, E.~L.~Cui, W.~Chen, X.~Liu, T.~G.~Steele and S.~L.~Zhu,
  {Eur.\ Phys.\ J.\ C {\bf 76}, 572 (2016)}.

\bibitem{Xiang:2017byz}
  J.~B.~Xiang, H.~X.~Chen, W.~Chen, X.~B.~Li, X.~Q.~Yao and S.~L.~Zhu,
  {Chin.\ Phys.\ C {\bf 43}, 034104 (2019)}.

\bibitem{Ioffe:1981kw}
  B.~L.~Ioffe,
  {Nucl.\ Phys.\ B {\bf 188}, 317 (1981)}
  Erratum: [{Nucl.\ Phys.\ B {\bf 191}, 591 (1981)}].

\bibitem{Ioffe:1982ce}
  B.~L.~Ioffe,
  {Z.\ Phys.\ C {\bf 18}, 67 (1983)}.

\bibitem{Espriu:1983hu}
  D.~Espriu, P.~Pascual and R.~Tarrach,
  {Nucl.\ Phys.\ B {\bf 214}, 285 (1983)}.

\bibitem{Yu:2017zst}
  F.~S.~Yu, H.~Y.~Jiang, R.~H.~Li, C.~D.~L{\"u}, W.~Wang and Z.~X.~Zhao,
  {Chin.\ Phys.\ C {\bf 42}, 051001 (2018)}.

\bibitem{Chen:2008qw}
  H.~X.~Chen, A.~Hosaka and S.~L.~Zhu,
  {Phys.\ Rev.\ D {\bf 78}, 054017 (2008)}.

\bibitem{Chen:2013jra}
  H.~X.~Chen,
  {Eur.\ Phys.\ J.\ C {\bf 73}, 2628 (2013)}.

\bibitem{Chen:2010ze}
  W.~Chen and S.~L.~Zhu,
  {Phys.\ Rev.\ D {\bf 83}, 034010 (2011)}.

\bibitem{Veliev:2010gb}
  E.~V.~Veliev, H.~Sundu, K.~Azizi and M.~Bayar,
  {Phys.\ Rev.\ D {\bf 82}, 056012 (2010)}.

\bibitem{Becirevic:2013bsa}
  D.~Be{\v c}irevi{\'c}, G.~Duplan{\v c}i{\'c}, B.~Klajn, B.~Meli{\'c} and F.~Sanfilippo,
  {Nucl.\ Phys.\ B {\bf 883}, 306 (2014)}.

\bibitem{Novikov:1977dq}
  V.~A.~Novikov, L.~B.~Okun, M.~A.~Shifman, A.~I.~Vainshtein, M.~B.~Voloshin and V.~I.~Zakharov,
  {Phys.\ Rept.\  {\bf 41}, 1 (1978)}.

\bibitem{Narison:2015nxh}
  S.~Narison,
  {Nucl.\ Part.\ Phys.\ Proc.\  {\bf 270-272}, 143 (2016)}.

\bibitem{Chang:2018aut}
  Q.~Chang, X.~N.~Li, X.~Q.~Li and F.~Su,
  {Chin.\ Phys.\ C {\bf 42}, 073102 (2018)}.

\bibitem{Liu:2007fg}
  X.~Liu, H.~X.~Chen, Y.~R.~Liu, A.~Hosaka and S.~L.~Zhu,
  {Phys.\ Rev.\ D {\bf 77}, 014031 (2008)}.

\bibitem{Chen:2017sci}
  H.~X.~Chen, Q.~Mao, W.~Chen, A.~Hosaka, X.~Liu and S.~L.~Zhu,
  {Phys.\ Rev.\ D {\bf 95}, 094008 (2017)}.

\bibitem{Cui:2019dzj}
  E.~L.~Cui, H.~M.~Yang, H.~X.~Chen and A.~Hosaka,
  {Phys.\ Rev.\ D {\bf 99}, 094021 (2019)}.

\bibitem{Shifman:1978bx}
  M.~A.~Shifman, A.~I.~Vainshtein and V.~I.~Zakharov,
  {Nucl.\ Phys.\ B {\bf 147}, 385 (1979)}.

\bibitem{Reinders:1984sr}
  L.~J.~Reinders, H.~Rubinstein and S.~Yazaki,
  {Phys.\ Rept.\  {\bf 127}, 1 (1985)}.

\bibitem{Grinstein:1990mj}
  B.~Grinstein,
  {Nucl.\ Phys.\ B {\bf 339}, 253 (1990)}.

\bibitem{Eichten:1989zv}
  E.~Eichten and B.~R.~Hill,
  {Phys.\ Lett.\ B {\bf 234}, 511 (1990)}.

\bibitem{Falk:1990yz}
  A.~F.~Falk, H.~Georgi, B.~Grinstein and M.~B.~Wise,
  {Nucl.\ Phys.\ B {\bf 343}, 1 (1990)}.

\bibitem{Dmitrasinovic}
  V.~Dmitra\v sinovi\' c and H.~X.~Chen,
  Phys. Rev. D \textbf{101}, no.11, 114016 (2020).

\bibitem{groundbaryon}
  {A full analysis on decay constants of charmed baryon fields}, in preparation.

\bibitem{Chen:2008qv}
  H.~X.~Chen, V.~Dmitrasinovic, A.~Hosaka, K.~Nagata and S.~L.~Zhu,
  {Phys.\ Rev.\ D {\bf 78}, 054021 (2008)}.

\bibitem{Chen:2009sf}
  H.~X.~Chen, V.~Dmitrasinovic and A.~Hosaka,
  {Phys.\ Rev.\ D {\bf 81}, 054002 (2010)}.

\bibitem{Chen:2010ba}
  H.~X.~Chen, V.~Dmitrasinovic and A.~Hosaka,
  {Phys.\ Rev.\ D {\bf 83}, 014015 (2011)}.

\bibitem{Chen:2011rh}
  H.~X.~Chen, V.~Dmitrasinovic and A.~Hosaka,
  {Phys.\ Rev.\ C {\bf 85}, 055205 (2012)}.

\bibitem{Dmitrasinovic:2016hup}
  V.~Dmitrasinovic, H.~X.~Chen and A.~Hosaka,
  {Phys.\ Rev.\ C {\bf 93}, 065208 (2016)}.

\bibitem{Chen:2012ex}
  H.~X.~Chen,
  {Eur.\ Phys.\ J.\ C {\bf 72}, 2180 (2012)}.

\bibitem{fierz}
  M.~Fierz, Z.~Physik, {\bf 104}, 553 (1937).

\bibitem{Chen:2006hy}
  H.~X.~Chen, A.~Hosaka and S.~L.~Zhu,
  {Phys.\ Rev.\ D {\bf 74}, 054001 (2006)}.

\bibitem{Chen:2006zh}
  H.~X.~Chen, A.~Hosaka and S.~L.~Zhu,
  {Phys.\ Lett.\ B {\bf 650}, 369 (2007)}.

\bibitem{Chen:2007xr}
  H.~X.~Chen, A.~Hosaka and S.~L.~Zhu,
  {Phys.\ Rev.\ D {\bf 76}, 094025 (2007)}.

\bibitem{Chen:2008ej}
  H.~X.~Chen, X.~Liu, A.~Hosaka and S.~L.~Zhu,
  {Phys.\ Rev.\ D {\bf 78}, 034012 (2008)}.

\bibitem{Chen:2018kuu}
  H.~X.~Chen, C.~P.~Shen and S.~L.~Zhu,
  {Phys.\ Rev.\ D {\bf 98}, 014011 (2018)}.

\bibitem{Beneke:1999br}
  M.~Beneke, G.~Buchalla, M.~Neubert and C.~T.~Sachrajda,
  {Phys.\ Rev.\ Lett.\  {\bf 83}, 1914 (1999)}.

\bibitem{Beneke:2000ry}
  M.~Beneke, G.~Buchalla, M.~Neubert and C.~T.~Sachrajda,
  {Nucl.\ Phys.\ B {\bf 591}, 313 (2000)}.

\bibitem{Beneke:2001ev}
  M.~Beneke, G.~Buchalla, M.~Neubert and C.~T.~Sachrajda,
  {Nucl.\ Phys.\ B {\bf 606}, 245 (2001)}.

\bibitem{Li:2020rcg}
  H.~D.~Li, C.~D.~L{\"u}, C.~Wang, Y.~M.~Wang and Y.~B.~Wei,
  JHEP {\bf 2004}, 023 (2020).

\bibitem{Landau}
  L.~D.~Landau and E.~M.~Lifshitz,
  {Quantum Mechanics (Non-Relativistic Theory)}, Pergamon Press, Oxford, 1977.




\end{thebibliography}
\end{document}